\documentclass[prc,superscriptaddress,amsmath,amssymb,showpacs,twocolumn]{
revtex4-1}

\usepackage{amssymb}
\usepackage{color}
\usepackage{graphicx}

\newcommand{\DSO}{{\ensuremath{\Delta_{\text{SO}}}}}
\newcommand{\DKK}{{\ensuremath{\Delta_{\text{KK'}}}}}
\newcommand{\vsd}{\ensuremath{V_{\text{sd}}}}
\newcommand{\vg}{\ensuremath{V_{\text{g}}}}

\newcommand{\un}[1]{{\ensuremath{\,\text{#1}}}}

\newcommand{\TK}{\ensuremath{T_{\text{K}}}}
\newcommand{\kB}{\ensuremath{k_{\text{B}}}}
\newcommand{\kTK}{\ensuremath{\kB\TK}}
\newcommand{\muB}{\ensuremath{\mu_{\text{B}}}}
\newcommand{\nel}{\ensuremath{N_{\text{el}}}}
\newcommand{\Bpar}{\ensuremath{B_\parallel}}
\newcommand{\Bperp}{\ensuremath{B_\perp}}
\newcommand{\SUF}{\ensuremath{SU(4)}}
\newcommand{\SUT}{\ensuremath{SU(2)}}
\newcommand{\Ecrit}{\ensuremath{\varepsilon_{\text{c}}}}

\DeclareMathAlphabet{\mathbbmsl}{U}{bbm}{m}{sl}

\begin{document}

\title{Broken SU(4) symmetry in a Kondo-correlated carbon nanotube}

\author{Daniel R. Schmid}
\affiliation{Institute for Exp. and Applied Physics, University of
Regensburg, 93040 Regensburg, Germany}

\author{Sergey Smirnov}
\affiliation{Institute for Theoretical Physics, University of
Regensburg, 93040 Regensburg, Germany}

\author{Magdalena Marga\'nska}
\affiliation{Institute for Theoretical Physics, University of
Regensburg, 93040 Regensburg, Germany}

\author{Alois Dirnaichner}
\affiliation{Institute for Exp. and Applied Physics, University of
Regensburg, 93040 Regensburg, Germany}
\affiliation{Institute for Theoretical Physics, University of
Regensburg, 93040 Regensburg, Germany}

\author{Peter L. Stiller}
\affiliation{Institute for Exp. and Applied Physics, University of
Regensburg, 93040 Regensburg, Germany}

\author{Milena Grifoni}
\affiliation{Institute for Theoretical Physics, University of
Regensburg, 93040 Regensburg, Germany}

\author{Andreas K. H\"uttel}
\altaffiliation{E-mail: andreas.huettel@ur.de}
\affiliation{Institute for Exp. and Applied Physics, University of
Regensburg, 93040 Regensburg, Germany}

\author{Christoph Strunk}
\altaffiliation{E-mail: christoph.strunk@ur.de}
\affiliation{Institute for Exp. and Applied Physics, University of
Regensburg, 93040 Regensburg, Germany}

\date{\today}


\begin{abstract}
Understanding the interplay between many-body phenomena and non-equilibrium 
in systems with entangled spin and orbital degrees of freedom is a central 
objective in nano-electronics. We demonstrate that the combination of 
Coulomb interaction, spin-orbit coupling and valley mixing results in a 
particular selection of the inelastic virtual processes contributing to the 
Kondo resonance in carbon nanotubes at low temperatures. This effect is dictated 
by conjugation properties of the underlying carbon nanotube spectrum at zero and 
finite magnetic field. Our measurements on a clean carbon nanotube are 
complemented by calculations based on a new approach to the non-equilibrium 
Kondo problem which well reproduces the rich experimental observations in Kondo 
transport.
\end{abstract}

\maketitle

\section{Introduction}
The Kondo effect \cite{Kondo1964} is an archetypical manifestation of strong
electronic correlations in mesoscopic systems. While first observed in bulk
metals with ferromagnetic impurities, it was shown to lead to a distinct 
zero-bias anomaly in the differential conductance of semiconductor quantum 
dots with odd electronic occupation \cite{GoldhaberGordon1998, 
GoldhaberGordon1998PRL, Kouwenhoven1998}. A degeneracy of quantum states 
required for its occurrence is usually provided by the electronic spin degree 
of freedom, resulting in the so-called \SUT\ Kondo behavior. Remarkably, in the 
Kondo regime the differential conductance obeys universal scaling as a function 
of temperature \cite{GoldhaberGordon1998PRL}, bias voltage 
\cite{prl-grobis:246601}, and magnetic field \cite{splitkondo}.

Clean carbon nanotubes (CNTs) \cite{nmat-cao:745} provide a unique test-bed for 
the investigation and manipulation of the quantum dot level structure and its 
consequences for the Kondo effect \cite{Nygaard2000}. In CNTs an additional 
degeneracy in the intrinsic low energy spectrum stems from two (K, K') graphene 
Dirac points and enables one to study unconventional correlation phenomena such 
as the orbital \SUT\ as well as spin plus orbital \SUF\ Kondo effects both 
experimentally \cite{nature-jarillo:484, Makarovski2007} and theoretically 
\cite{Borda2003, Choi2005, Lim2006,Fang2008, Fang_Erratum, Lim2011,  Galpin2010}.
Besides revealing the curvature induced spin-orbit interaction 
\cite{nature-kuemmeth:7186, prb-delvalle:165427, ncomms-steele:1573, Ando2000}, 
experiments indicate that also the K-K' degeneracy is frequently lifted with a 
finite energy \DKK\ \cite{nature-kuemmeth:7186, Jespersen2011, 
GroveRasmussen2012}. While usually attributed to the presence of disorder in 
damaged or contaminated CNTs, it is observed also in clean carbon nanotubes as 
a contribution from the CNT's contact interfaces \cite{Marganska2014}. 

In the following, we demonstrate how this type of \SUF\ symmetry breaking leads 
to unconventional Kondo transport phenomena. Our results clearly show that the 
Kondo behavior at zero magnetic field is controlled by time reversal symmetry, 
which allows to identify two distinct, two-fold degenerate Kramers doublets. The 
analysis of the breaking of this symmetry in a magnetic field parallel (\Bpar) 
or perpendicular (\Bperp) to the CNT axis leads to a detailed understanding of 
the many-body processes contributing to transport in the Kondo regime. It allows 
to disentangle the role of the conjugation relations and goes significantly 
beyond earlier studies on Kondo ions \cite{Garnier1997, Reiner2001, Ernst2011}, 
where Kondo satellites could be observed but the tunability of the spectrum by 
a magnetic field was not given. 

As discussed below, our studies also significantly advance the understanding of 
Kondo physics in CNTs in the deeply non perturbative regime. In particular, we 
elucidate the reasons for the absence of certain many-body transitions expected 
from previous theoretical works \cite{Fang2008, Fang_Erratum, Lim2011} but not 
observed in our as well as in previous experiments \cite{Quay2007, Lan2012, 
Cleuziou2013}.

\section{Transport measurements}
Electronic transport measurements have been performed on a clean, freely 
suspended single-wall carbon nanotube (CNT) contacted with rhenium and
capacitively coupled to a global back gate at milli-Kelvin temperatures. The
carbon nanotube was grown by chemical vapor deposition across pre-defined
trenches and electrode structures to minimize damage and contamination
mechanisms \cite{nature-kong:878, nmat-cao:745}.
\begin{figure*}
\includegraphics[width=15cm]{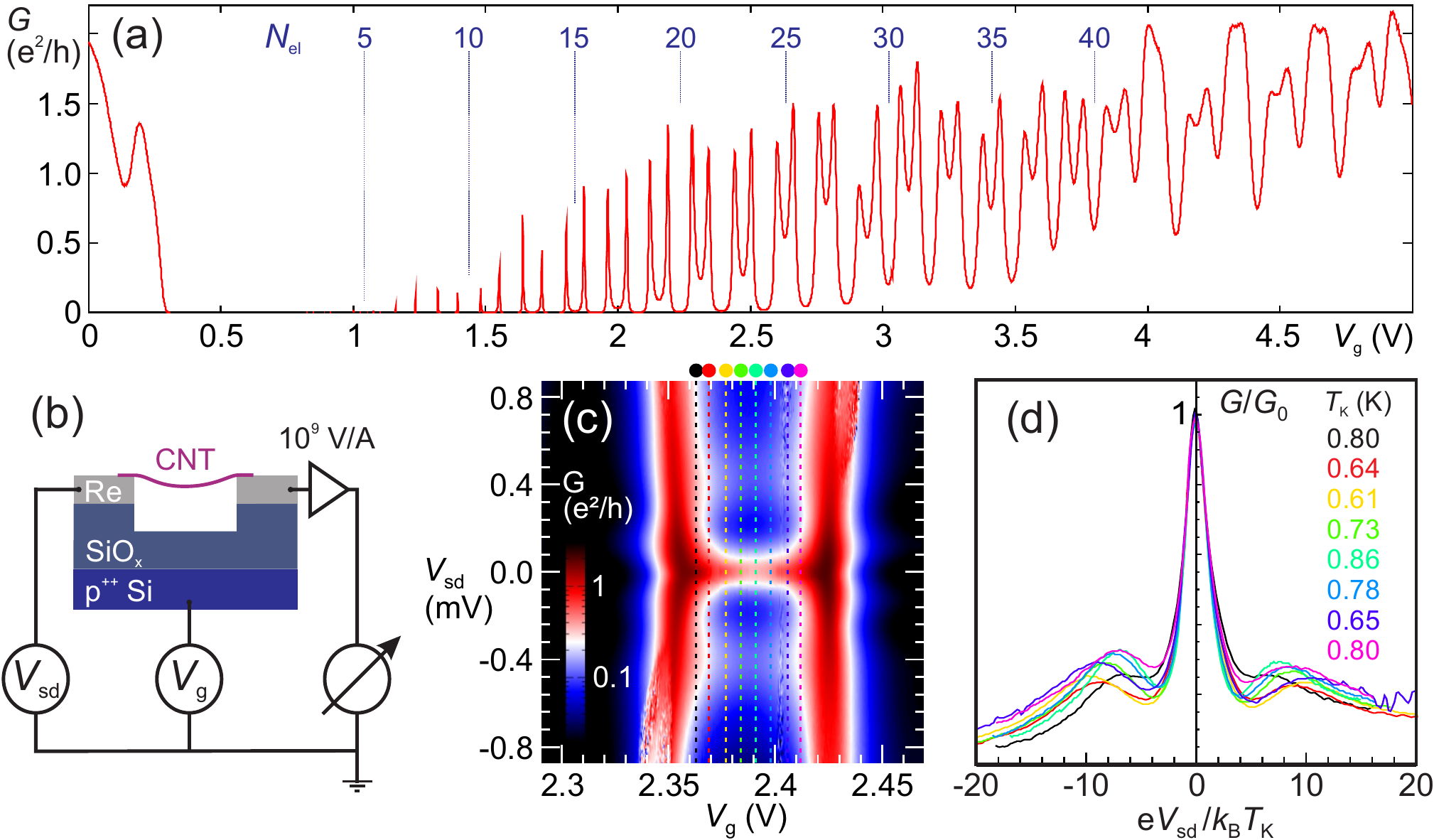}
\caption{\label{fig:ovbias} (Color) Measurement of the Kondo effect in 
transport through an ultra-clean carbon nanotube.
(a) Low-bias conductance ($\vsd=0.2\un{mV}$) through a clean, suspended small 
band gap carbon nanotube as function of applied gate voltage \vg. 
(b) Device geometry and electronic measurement setup for differential 
conductance measurement. 
(c) Differential conductance $G(\vg,\vsd) = \un{d}I(\vg,\vsd) / 
\un{d}V_{\text{sd}}$ inside and around the gate voltage window with $\nel=21$ 
($T=30\un{mK}$). A sharp Kondo ridge at zero bias voltage $\vsd=0$ and broader 
satellite ridges at finite bias $\vsd \simeq \pm 0.5 \un{mV}$ are clearly 
visible. (d) Line traces $G(\vsd)$ at constant \vg\ corresponding to colored 
dotted lines in (c), with \vsd\ rescaled by the corresponding Kondo temperature 
\TK\ (see text). The central conductance peak at $\vsd=0$ displays the universal 
Kondo behavior, while the satellite peaks move. A similar plot across different 
Coulomb oscillations can be found in the Appendix, Fig.~\ref{shellsrescaled}. 
}
\end{figure*}
As seen in the low-bias conductance (Fig.~\ref{fig:ovbias}(a)), a small band 
gap separates a Fabry-P\'{e}rot pattern \cite{nature-liang:665} in the highly 
transparent hole regime from sharp Coulomb blockade oscillations in the few 
electron regime ($\nel<10$). With increasing gate voltage enhanced conductance 
is observed, leading in particular to a Kondo zero-bias anomaly in the odd 
electron number valleys. The electronic setup used for the measurements is 
sketched in Figure~\ref{fig:ovbias}(b). A dc- and an ac-voltage are 
superimposed and applied as bias voltage $\vsd = \vsd^{\text{dc}} + 
\vsd^{\text{ac}}$ to the source contact. The current from the drain contact is 
converted to a voltage and measured with a lock-in amplifier. The highly 
positive doped silicon substrate acts as global back gate. 

In the following, we focus on the intermediate coupling regime and measure the
differential conductance as a function of gate voltage \vg\ and bias voltage
\vsd\ (Figure~\ref{fig:ovbias}(c)). Besides the pronounced conductance ridge
at zero bias voltage, additional broad satellite peaks appear symmetrically at
finite bias voltage $\vsd \simeq \pm 0.5 \un{mV}$, depending only weakly on 
the gate voltage. In analogy to the case of a broken spin degeneracy in a 
magnetic field, these satellite peaks at zero magnetic field signal a 
lifted degeneracy of the ground state, allowing inelastic transport processes 
to take place. Finite bias conductance peaks together with a zero bias Kondo 
peak have already been observed in CNT quantum dots with odd shell filling 
\cite{Nygaard2000, Quay2007, Lan2012, Cleuziou2013}. In the latter three 
experiments the evolution of the satellites in perpendicular \cite{Quay2007} and 
parallel \cite{Lan2012, Cleuziou2013} magnetic fields have been reported. 
Because a finite field breaks time reversal symmetry, inelastic transitions 
between Zeeman splitted or orbitally splitted levels are expected to become 
visible \cite{Jespersen2011, Fang2008}. Strikingly, in the three experiments 
\cite{Quay2007, Lan2012, Cleuziou2013} not all of the inelastic transitions 
expected from (possibly Kondo enhanced) cotunneling \cite{Jespersen2011} or for 
the non perturbative Kondo regime \cite{Fang2008} could be seen. In this work we 
clarify the nature of the transitions contributing to the finite bias peaks.

Non-equilibrium co-tunneling is a threshold effect which can  give rise to a 
step-like cusp in the differential conductance \cite{Wegewijs2001}. Kondo 
correlations treated within lowest order perturbation theory yield a logarithmic 
enhancement of this cusp which emphasizes further the threshold effect 
\cite{Paaske2006}. This behavior is expected in the perturbative regime $T > 
\TK$ of temperatures larger than the Kondo temperature. In the strong coupling 
regime $T < \TK$ a perturbative treatment of Kondo correlations is no longer 
appropriate. At such low temperatures, and for $k_{\rm B} T_K \sim  \Delta$, 
where $\Delta$ is the energy of the inelastic transition associated to valley 
mixing and spin-orbit coupling, true Kondo peaks at finite bias, rather than 
co-tunneling cusps, are expected to develop \cite{Fang2008, Galpin2010}. This is 
the parameter regime in \cite{Quay2007, Lan2012, Cleuziou2013} and, as 
demonstrated below, also of our experiments. Hence, the conductance traces in 
Figure~\ref{fig:ovbias}(d) are a manifestation of the Kondo effect in the strong 
coupling regime. As we shall show, the interplay of Coulomb interaction and the 
intrinsic symmetry properties of the CNT-Hamiltonian yields a selective 
enhancement of virtual processes contributing to the Kondo effect in the non 
perturbative regime. 

\begin{figure}
\includegraphics[width=\columnwidth]{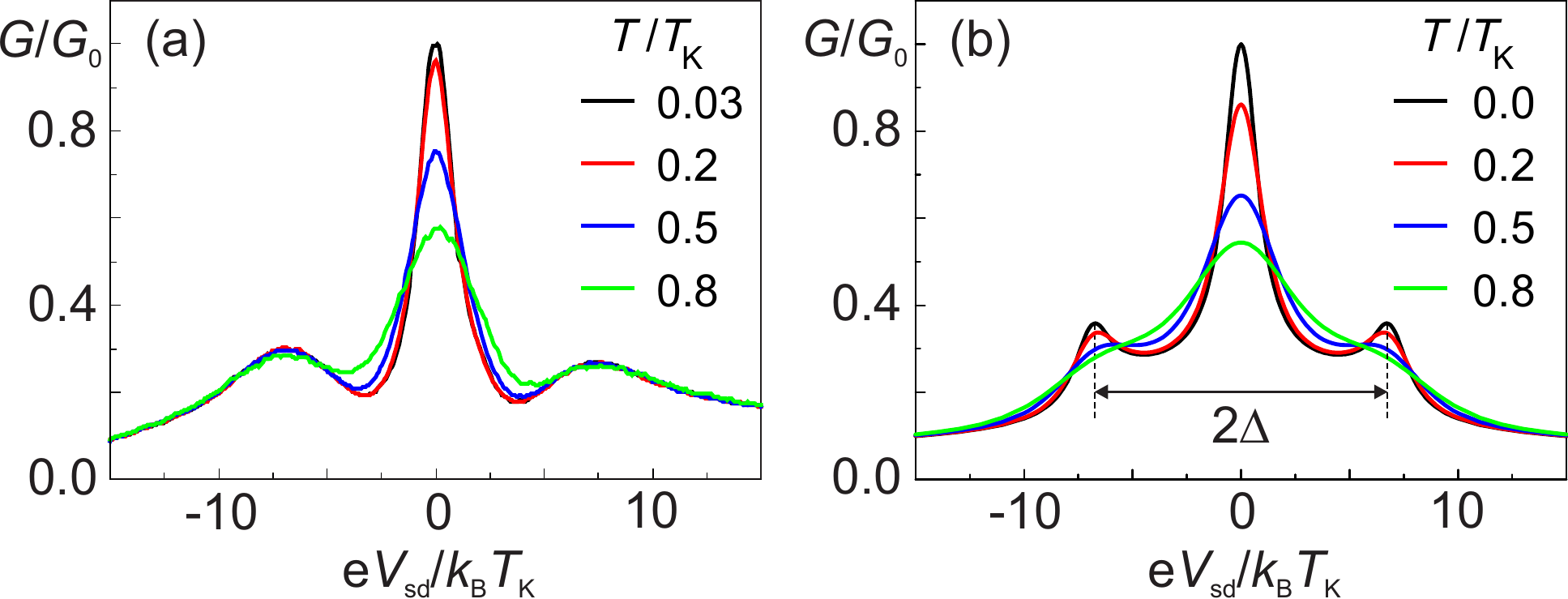}
\caption{\label{fig:tempdep} (Color online) Temperature dependence of the main 
Kondo resonance and its satellites.
(a) Measured differential conductance traces $G(\vsd)$ at different 
temperatures, normalized by $G_0=G(\vsd=0)$ for the lowest temperature at $\vg 
= 2.39\un{V}$; the bias voltage is scaled with the Kondo temperature $\TK = 
0.86\un{K}$. 
(b) Differential conductance obtained from our field-theoretical calculation. 
The distance between the satellite peaks at $T=0$ used in the calculation is 
$2\Delta=13.8\,\kTK$.
}
\end{figure}
\section{Universality}
To unambiguously claim that a zero-bias anomaly observed in experiments has
Kondo origin, the characteristic universal scaling behavior with the energy
scale determined by the Kondo temperature \TK\ has to be tested. We record
conductance traces $G(\vsd)$ at different discrete gate voltage values \vg\
within the $21^{\text{st}}$ Coulomb diamond. For each such trace, we determine
the Kondo temperature $\TK(\vg)$ in non-equilibrium from the central peak in 
the bias voltage trace using the condition $G(\kB\TK/e)\simeq 2G_0/3$ 
and $G_0=G(\vsd=0)$ \footnote{The precise expression for the Kondo temperature 
is extracted from the many-body theory calculations, yielding $G(k_B T_K / e) = 
0.612\,G_0$. Experimentally, fluctuations of $G(\vsd \simeq 0)$ 
between different measurement runs were observed, leading to corresponding 
fluctuations of the obtained \TK\ values. For consistency we use 
$\TK=0.86\un{K}$ throughout the evaluation of the center of valley $\nel=21$.} 
\cite{Kretinin_2012, Pletyukhov2012}. We then rescale the bias voltage with the 
respective Kondo temperature, and normalize the conductance to its maximum 
value $G_0\sim 0.5 e^2/h$. The collapse of all curves $G(e\vsd/\kB\TK)/G_0$ 
around $\vsd=0$ into universal behavior, as illustrated in 
Fig.~\ref{fig:ovbias}(d), clearly demonstrates the Kondo origin of the 
zero-bias feature. This behavior can be compared with the theoretical curves in 
Fig.~\ref{figure_3_supp}, where we show that the universal line shape of the 
central Kondo resonance remains essentially unchanged also in the transition 
between \SUT\ and \SUF.

\begin{figure}
\includegraphics[width=\columnwidth]{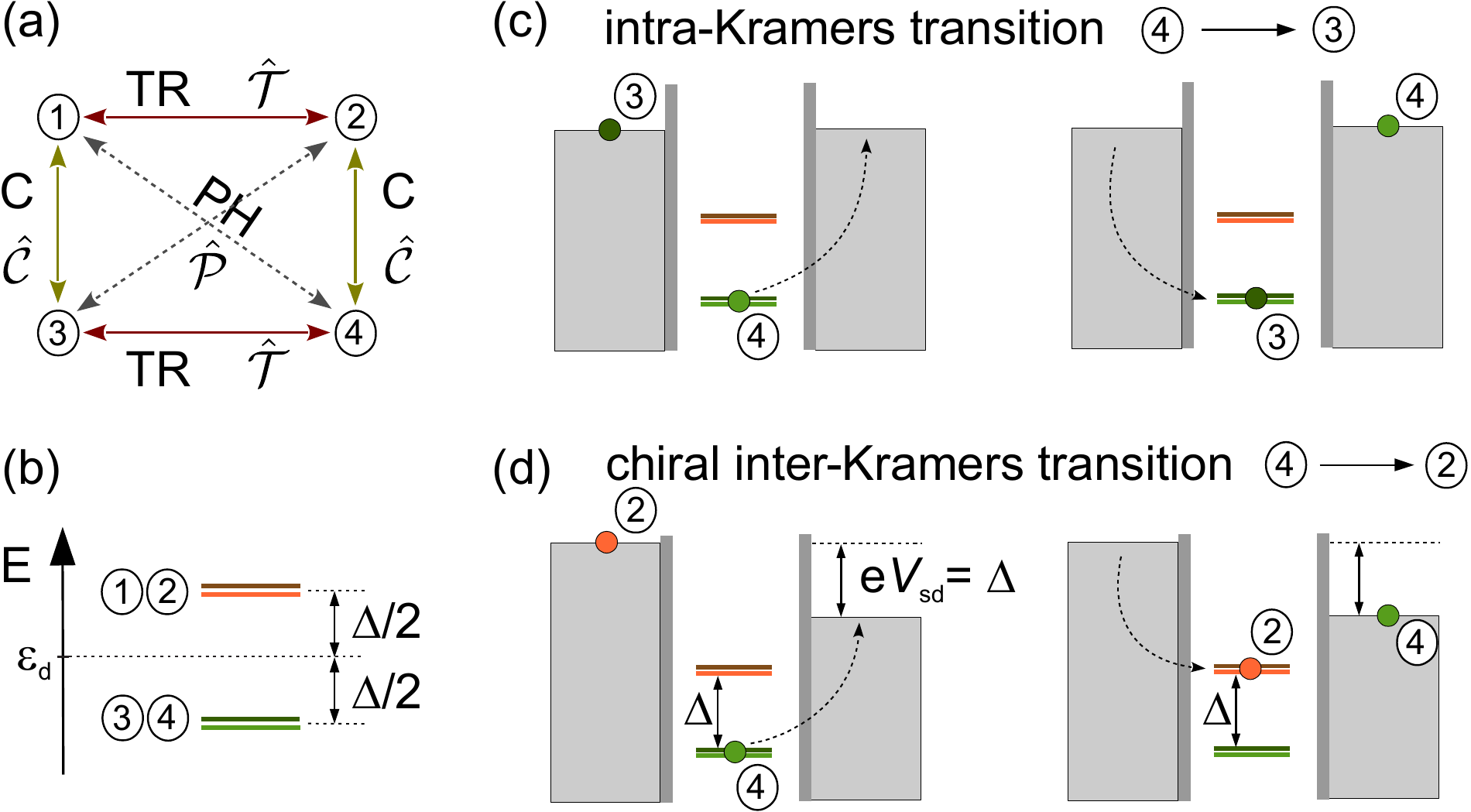}
\caption{\label{fig:symmetries}  (Color online) Conjugation relations, level 
spectrum, and selected Kondo transport processes.
(a) Energy levels associated with a longitudinal mode of a CNT accounting for 
spin and valley degrees of freedom. The time-reversal operator 
$\hat{\mathcal{T}}$ connects the states $(1,2)$ and $(3,4)$. The operators 
$\hat{\mathcal{C}}$ and $\hat{\mathcal{P}}$ govern chiral and particle-hole 
conjugation and provide further pairs of conjugated states. Chiral pairs are 
$(1,3)$ and $(2,4)$; particle-hole pairs are $(1,4)$ and $(2,3)$. 
(b) Spin-orbit coupling and valley mixing break the four-fold degeneracy but 
not the time-reversal symmetry. The spectrum splits in two degenerate Kramers 
doublets $(1,2)$ and $(3,4)$, respectively, separated by the energy difference 
$\Delta=\sqrt{\DKK^2+\DSO^2}$. 
(c) The Kondo peak at zero bias is governed by virtual processes within the 
Kramers pairs $(1,2)$ and $(3,4)$ (intra-Kramers transitions). 
(d) The satellite peaks at finite bias, $\vsd=\pm \Delta/e$, result from 
inelastic transitions within the chiral pairs, $(1,3)$ and $(2,4)$ 
(chiral inter-Kramers transitions).}
\end{figure}

After the rescaling the position of the satellite peaks varies, i.e., here the 
universality is apparently lost. This is to be expected since the Kondo 
temperature \TK\ varies within the Coulomb valley region but the splitting 
$\Delta$ between central peak and satellites is gate independent. Complete 
universality of the differential conductance, i.e., universality in the whole 
range of voltages, requires the ratio $\Delta/\kTK$ to be invariant 
\cite{Yamada1984}. In general one obtains $\TK(\Delta) = \TK(0) f[\Delta/k_{\rm 
B}\TK(0)]$, where $\TK(0)=\TK^{SU(4)}$ is the Kondo temperature for the \SUF\ 
Kondo effect, and $f(x)$ depends on the strength of the \SUF\ symmetry 
breaking. In our experiment we find $\Delta/\kTK\simeq 7$. As shown in 
Ref.~\cite{Galpin2010}, this implies that in our experiment the \SUF\ symmetry 
is weakly broken.

Finally, we notice a nonmonotonic dependence of \TK\ on the gate voltage \vg, 
with a local maximum in the center of the Coulomb blockade region. Such 
behavior is not expected for the \SUT\ Anderson model, which rather predicts a 
local minimum \cite{Schrieffer1966}. Hence, the peculiar voltage dependence 
observed in our experiment may as well be a signature of weakly broken \SUF.  

\section{Temperature dependence}
Figure~\ref{fig:tempdep}(a) displays the temperature dependence of $G(\vsd, T)$ 
in the center of the same Coulomb diamond ($\nel=21$), where $\vg = 
2.39\,\un{V}$. The central peak behaves in a way characteristic for the Kondo 
effect: it is suppressed and broadened for increasing temperatures. The 
satellite peaks are increasingly washed out at elevated temperatures.  A slight 
bias asymmetry is observed in the curves, which we attribute to asymmetries in 
the couplings to the leads. Such asymmetries are also responsible for the 
reduction of the maximum $G_0$ with respect to the  unitary limit value $2e^2/h$ 
expected for a fully symmetric set-up.

Figure~\ref{fig:tempdep}(b) displays the differential conductance obtained from 
our calculation based on the slave boson Keldysh effective action formalism 
\cite{Altland_2010} discussed in Sec.~\ref{sec:modeling} and in the Appendix. 
The calculation uses the minimal model Hamiltonian Eq. (\ref{Ham_zero_B}) for a 
single 
longitudinal mode of a CNT including spin orbit interaction (with the energy 
scale \DSO) and valley mixing (with characteristic energy scale \DKK). These 
couplings break the orbital degeneracy of the CNT spectrum (and hence the $\SUF$ 
symmetry) but preserve time-reversal symmetry. As a result, the non interacting 
CNT spectrum displays two degenerate Kramers doublets separated by the spacing 
$\Delta=\sqrt{\DKK^2+\DSO^2}$. The simulation uses the value of $2\Delta=13.8 
k_{\rm B}\TK$ obtained from the experiment. For simplicity, and to stress the 
universal features of the problem, a symmetric coupling to the leads has been 
used in the simulation, as well as equal coupling of the CNT modes to the leads. 
Hence, in our calculation the central peak  reaches at zero temperature the 
unitary limit $2e^2/h$. To compare with the experiment, the theoretical curves 
have been normalized by the maximum theoretical value $G_0=2e^2/h$. Our 
calculation reproduces well the experimentally observed evolution of peak 
amplitudes with temperature. The tails at high voltages decay faster than in the 
experiment. This behavior is due to our approximation scheme for the Keldysh 
effective action, where only terms quadratic in the slave bosonic fields are 
retained. Within this approximation, the behavior of the central  and inelastic 
peaks has been proven to be accurately reproduced for the $\SUT$  Anderson model 
\cite{Smirnov_2013, Smirnov_2013a}. To improve the description of the tails 
quartic terms should be included. Such treatment, however, would go beyond the 
scope of this work. 

\section{Conjugation relations}
\label{sec:conjugation}
A quantitative analysis of our experimental results has to combine the 
properties of the underlying set of single-particle states with the 
Kondo-correlations. In this section we analyze conjugation relations valid for 
the single particle spectrum which turn out to have significant impact on the 
many-body properties of our CNT quantum dot. In the simplest model for a CNT one 
expects a four-fold degenerate longitudinal level at the energy $\varepsilon_d$ 
\cite{Laird2014}. This degeneracy is removed by KK' valley mixing and spin-orbit 
interaction with splitting $\Delta$. We denote the four resulting energies 
associated to the eigenstates $|1\rangle,|2\rangle,|3\rangle,|4\rangle$ by 
$\varepsilon_1, \varepsilon_2,\varepsilon_3,\varepsilon_4$, respectively. At 
zero magnetic field, the effective CNT-Hamiltonian $\hat{H}^{(0)}_{\text{CNT}}$ 
(see Eq.~(\ref{Ham_zero_B}) in the Appendix) displays time-reversal (TR) 
symmetry governed by the operator $\hat{\mathcal{T}}$. The 
${\mathcal{T}}$-conjugated pairs of states are the Kramers doublets $(1,2)$ and 
$(3,4)$, with $\varepsilon_1 = \varepsilon_2$, and $\varepsilon_3 = 
\varepsilon_4 = \varepsilon_1-\Delta$, see Fig.~\ref{fig:symmetries}(a). 
Additional operators $\hat{\mathcal{P}}$ and $\hat{\mathcal{C}}$ can be 
introduced, which anticommute with $\hat{H}^{(0)}_{\text{CNT}}$ and which allow 
to connect the states within one quadruplet in the way depicted in 
Fig.~\ref{fig:symmetries}(a) (see Appendix~\ref{app:singleparticle} for further 
details). We call the operations related to $\hat{\mathcal{P}}$ and 
$\hat{\mathcal{C}}$ particle-hole (PH) \cite{Altland1997, footnote_PHS} and 
chiral (C) conjugation, respectively. The operator $\hat{\mathcal{P}}$ 
conjugates states from different Kramers doublets; the pairs are $(1,4)$ and 
$(2,3)$ with $\varepsilon_1 (\Delta)=\varepsilon_d +\Delta/2 = \varepsilon_4 
(-\Delta) $, and $\varepsilon_2(\Delta)=\varepsilon_d +\Delta/2 
=\varepsilon_3(-\Delta)$, as displayed in Fig.~\ref{fig:symmetries}(b). If TR 
and PH conjugation hold, so does chiral conjugation, which is represented by 
the operator $\hat{\mathcal{C}} = \hat{\mathcal{P}}\hat{\mathcal{T}}^{-1}$. The 
chirally conjugated pairs are $(1,3)$ and $(2,4)$. 
\begin{figure}
\begin{center}
\includegraphics[width=\columnwidth]{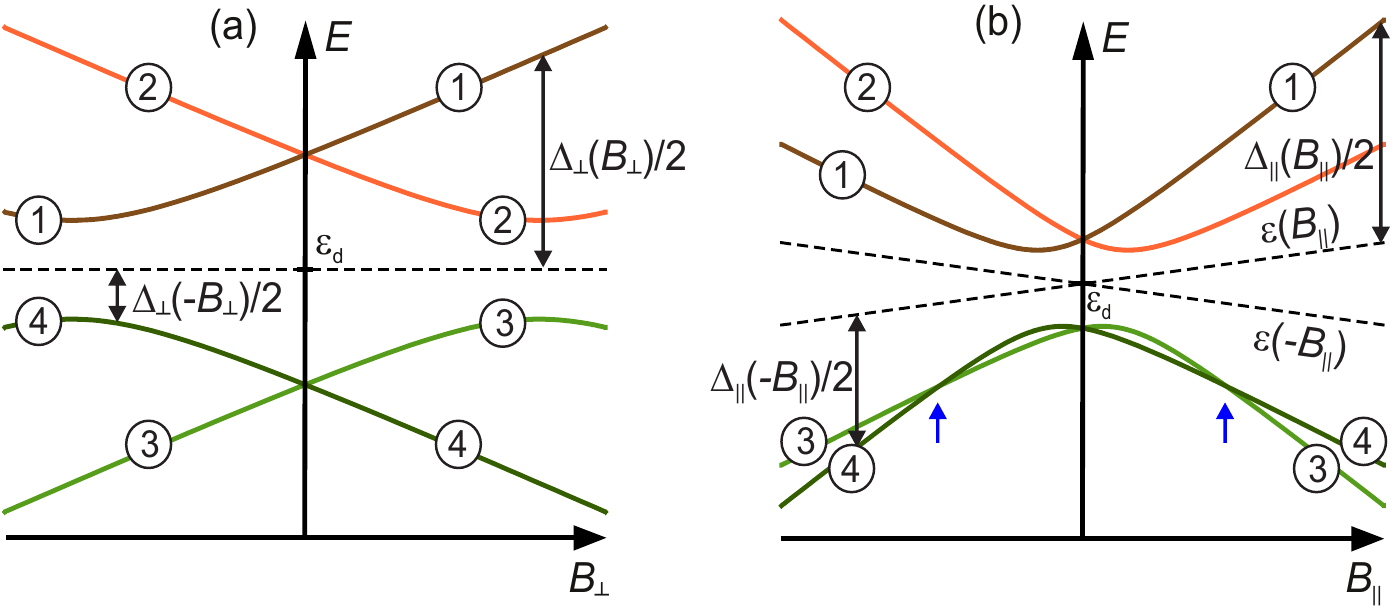}
\end{center}
\caption{\label{fig:elevels} (Color online) Sketch of the evolution of the 
energy levels in a magnetic field (a) perpendicular and (b) parallel to the CNT 
axis  in the parameter regime $\DKK>\DSO$. The energy eigenstates are labeled 
$1-4$. At $B=0$ there are two energy degenerate Kramers doublets $(1,2)$ and 
$(3,4)$. At finite fields the degeneracy is lifted and the conjugation 
relations $\varepsilon_1(\vec{B}) = \varepsilon_2(-\vec{B})$ and 
$\varepsilon_3(\vec{B})=\varepsilon_4(-\vec{B})$ hold. 
(a) In perpendicular field the energy difference within each chiral pair, i.e., 
$(1,3)$ and $(2,4)$, remains essentially independent of \Bperp, provided that 
$\Bperp \lesssim \DKK/g_{\text{s}}\muB$.  Moreover,  the 
particle-hole pairs $(1,4)$ and $(2,3)$ are related by the conjugation
relations $\varepsilon_1(\Delta_\perp(\Bperp)) =\varepsilon_d  + 
\Delta_\perp(\Bperp)/2 = \varepsilon_4(-\Delta_\perp(\Bperp))$ and 
$\varepsilon_2(\Delta_\perp(-\Bperp))=\varepsilon_d+
\Delta_\perp(-\Bperp)/2=\varepsilon_3(-\Delta_\perp(-\Bperp))$. 
b) For \Bpar\ the Aharonov-Bohm effect induces a level crossing of the pair 
$(3,4)$, as indicated by the blue arrows. Again  $\mathcal{P}$-conjugation 
holds. E.g., for the pair $(1,4)$  we find the relation 
$\varepsilon_1(\Bpar,\Delta_\parallel(\Bpar))=\varepsilon(\Bpar)+
\Delta_\parallel(\Bpar)/2 = \varepsilon_4(\Bpar,-\Delta_\parallel(\Bpar))$.
}
\end{figure}

The zero-bias Kondo peak is necessarily induced by transitions between the
degenerate states of the time-reversed Kramers pairs $(1,2)$ and $(3,4)$, which 
we call 'intra-Kramers' transitions, see Fig.~\ref{fig:symmetries}(c). A similar 
reasoning applies for the finite bias Kondo peaks at voltages equal to 
$\pm\Delta/e$: the inelastic peaks are necessarily induced by transitions 
between distinct Kramers pairs, called in the following 'inter-Kramers 
transitions. From Figs. ~\ref{fig:tempdep}(a) and (b) no further information on 
the nature of the inelastic transitions can be extracted. 

Additional insight can be obtained by looking at the evolution of the central 
peak and of its satellites in finite magnetic fields, as TR is broken and hence 
Kramers degeneracy is lifted. Figure~\ref{fig:elevels}(a) and 
\ref{fig:elevels}(b) display the dispersion of the four single particle states 
in a magnetic field perpendicular and parallel to the tube axis. It is clearly 
visible that in a finite magnetic field, $\vec{B}$, conjugation relations 
persist that lead to close connections between the single particle energies, 
e.g. $\hat{\mathcal{T}}|1,\vec{B}\rangle = |2,-\vec{B}\rangle $ leads to  
$\varepsilon_1(\vec{B}) =\varepsilon_2(-\vec{B})$, and  $\hat{\mathcal{P}} 
|1,\vec{B}\rangle = |4,\vec{B}\rangle $ leads to 
$\varepsilon_1(\vec{B},\Delta(\vec{B}))=  
\varepsilon_4(\vec{B},-\Delta(\vec{B}))$, with $\Delta(\vec{B})$ being a 
magnetic field dependent level splitting (see Eqs. (\ref{Eig_en_perp}), 
(\ref{Delta_perp}) and (\ref{Energies_perp}) of Appendix 
\ref{app:singleparticle}).

\begin{figure*}
\includegraphics[width=15cm]{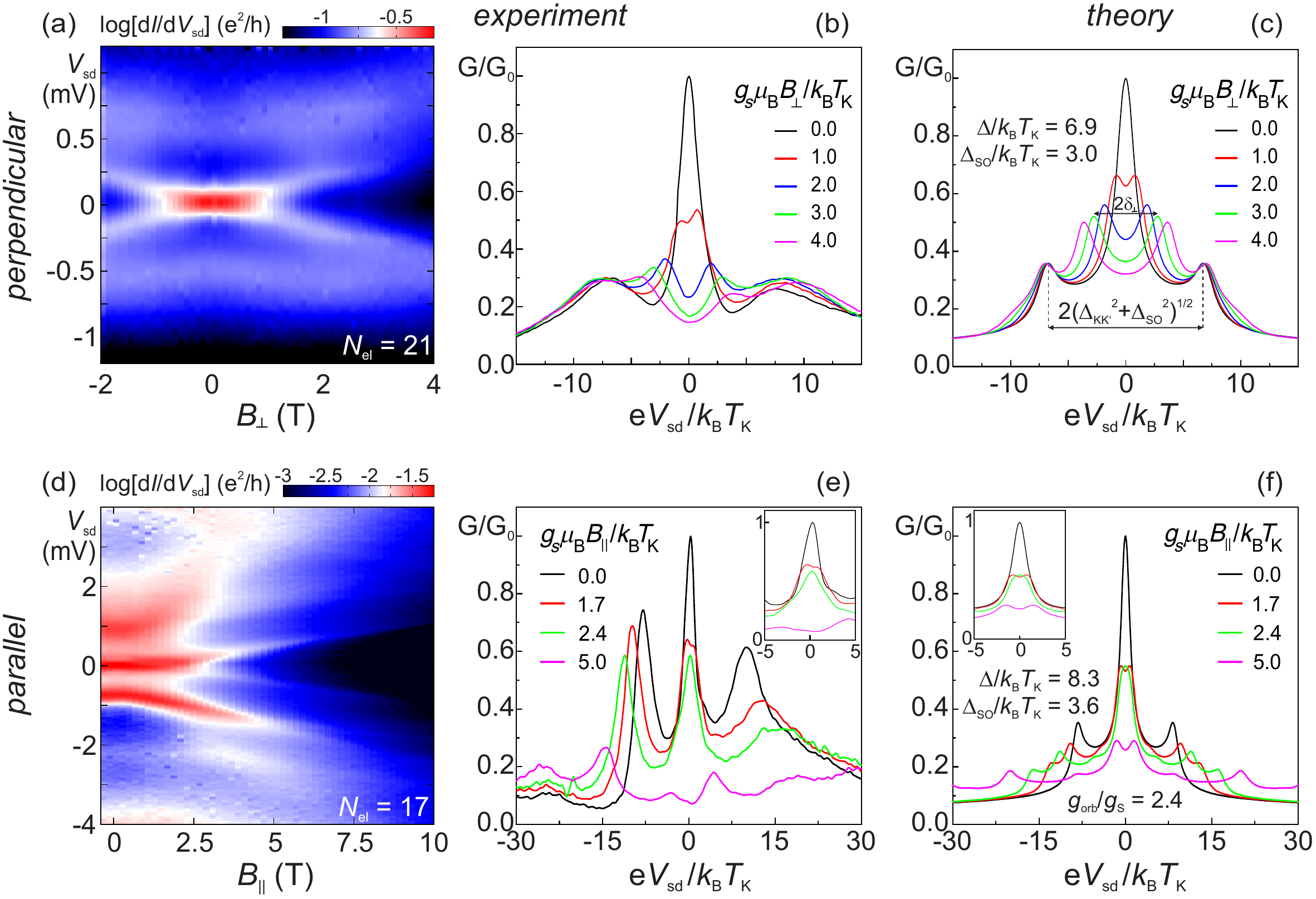}
\caption{\label{fig:magnetic} (Color) Kondo differential conductance in a 
magnetic field.
(a) Measured differential conductance $G(\Bperp,\vsd)$ as a function of the 
bias voltage \vsd\ and magnetic field \Bperp\ perpendicular to the CNT axis at
$\vg=2.39\,\un{V}$ corresponding to the electron number $\nel=21$. The central 
peak splits at $\Bperp>0.7\,\kTK/g_{\text{s}}\muB$, while the satellites 
remain nearly unaffected.
(b) Line traces $G(\vsd)$ versus bias voltage (rescaled with $\TK=0.86\un{K}$) 
from (a) for several values of \Bperp. 
(c) Theoretical results for the differential conductance at zero temperature 
for corresponding values of $g_{\text{s}}\muB\Bperp/\kTK$ showing good 
agreement with the experiment \cite{note:G-asymm}.
(d) Measurement as in (a) but in parallel alignment of the magnetic field, 
$G(\Bpar,\vsd)$, $\nel=17$. 
(e) Line traces from (d), where \vsd\ is scaled with the Kondo temperature
$\TK=1.12\un{K}$ and the conductance is scaled using its maximum $G_0 = 
G(\vsd=0)$ at $\Bpar=0\un{T}$. 
(f) Theoretical results for the differential conductance at zero temperature 
for different values of \Bpar. The insets in (e) and (f) show the evolution of 
the central peak in the experiment and theory, respectively.}
\end{figure*}
As revealed by the evolution of the satellite peaks in magnetic field shown in 
the next section, they only involve transitions among the chiral pairs $(1,3)$ 
and $(2,4)$ [see Fig.~\ref{fig:symmetries}(d)], while transitions between 
${\mathcal{P}}$-conjugated states are absent. Our observations seem to be 
consistent with other data in the non perturbative regime shown in 
\cite{Quay2007, Lan2012, Cleuziou2013}. Also in those experiments no 
${\mathcal{P}}$-transitions could be resolved in finite magnetic fields. For 
example, in the experiment by Cleuziou et al., Ref. \cite{Cleuziou2013}, only 
one of the two expected excitation lines (called $\beta$ and $\delta$ by the 
authors) could be identified, see Fig. 2(a) in Ref. \cite{Cleuziou2013}. Using 
the parameters given in that paper, we find that the non-observed inelastic 
transition is $\delta$, corresponding in our terminology to a 
$\mathcal{P}$-transition.

\section{Evolution of the Kondo peaks in magnetic field}
\begin{figure}
\includegraphics[width=0.75\columnwidth]{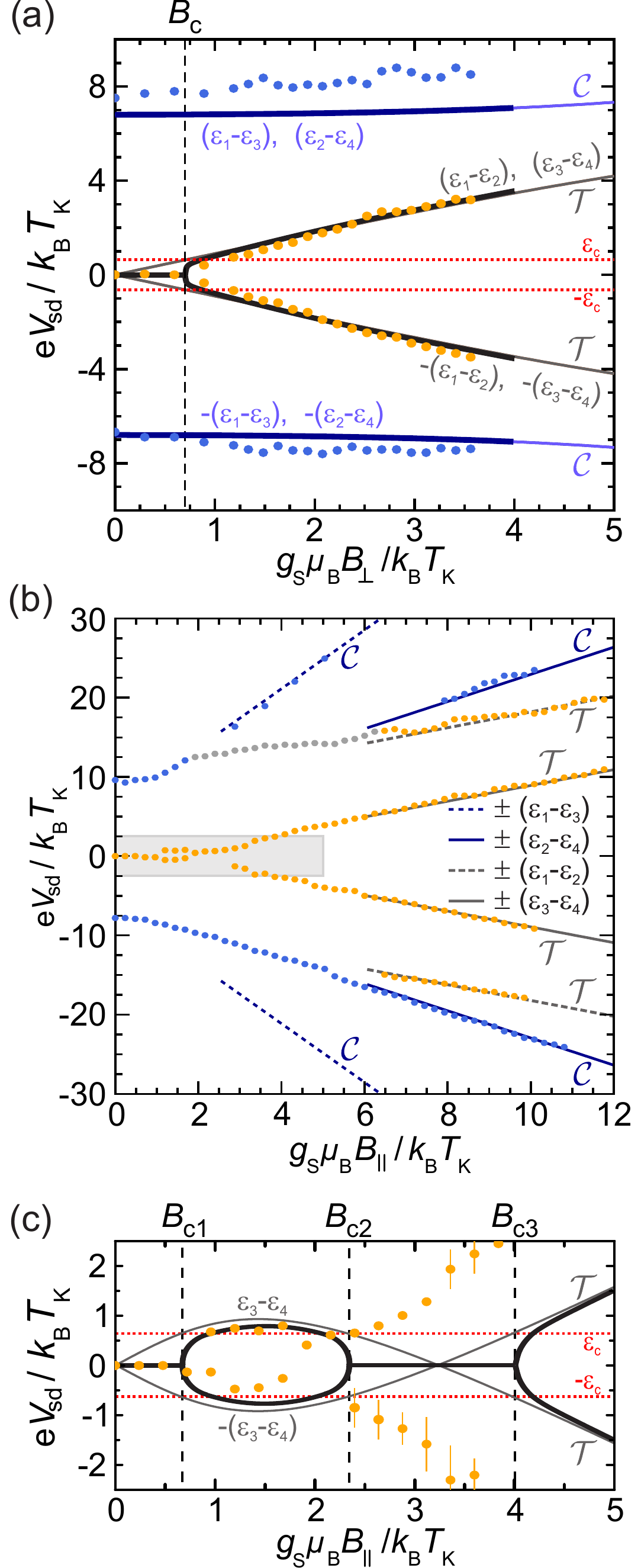}
\caption{\label{fig:comparison} (Color) Magnetic field behavior of the Kondo 
peaks.
(a) Predicted evolution of the Kondo peaks in perpendicular field taking into 
account many-body effects (thick solid lines) together with single particle 
energy differences (thin solid lines) and the experimental data (dots). A 
splitting occurs when the energy difference within the lowest Kramers doublet 
$|\varepsilon_3-\varepsilon_4|$ exceeds $\Ecrit$. 
(b) Experimental  evolution of the Kondo peaks in parallel magnetic field  
(dots), compared with  single-particle energy differences (solid and dashed 
lines) evaluated at large fields. The grey rectangle identifies the low field 
and low bias region analyzed in panel (c). 
(c) Evolution of the low-bias Kondo peaks in parallel magnetic field. In this 
case $\delta_{\parallel,1}=|\varepsilon_3-\varepsilon_4|$ (grey solid line) 
first grows with increasing field until it exceeds the threshold $\Ecrit$ (red 
dotted line) at the critical field $B_{\text{c1}}$, beyond which the peak 
starts to split (see dots and thick solid line). The predicted peak splitting 
vanishes again at $B_{c2}$. Above $B_{c3}$ it grows again, first in a sublinear 
way and afterwards linearly with the magnetic field.}
\end{figure}

The behavior of the Kondo peaks in magnetic field, reported in 
Figs.~\ref{fig:magnetic} and \ref{fig:comparison}, provides a sensitive tool 
that allows us to discriminate between the different types of Kondo-enhanced 
transitions. In fact, the positions of the Kondo peaks are related to the 
energy differences between the two dot states involved in the transition, and 
these depend very differently on direction and strength of the magnetic field 
for transitions between ${\mathcal{T}}$, ${\mathcal{C}}$, or ${\mathcal{P}}$ 
pairs. The central Kondo peak results from intra-Kramers transitions and its 
splitting reveals the breaking of time reversal symmetry. From the Keldysh 
effective action theory (see Sec.~\ref{sec:modeling}) a splitting of the 
central Kondo resonance is expected once the energetic separation within a 
Kramers doublet exceeds a threshold value $\Ecrit$, as observed in 
Figs.~\ref{fig:magnetic}(a),(b) and \ref{fig:magnetic}(d),(e). 

In perpendicular fields puzzling at first glance is the independence of the 
positions of the satellite Kondo peaks on the field  
[Figs.~\ref{fig:magnetic}(a),(b)]. This is in strong contrast to the 
co-tunneling regime investigated earlier (see Fig. 3 in 
Ref.~\onlinecite{Jespersen2011}), where a splitting of the inelastic 
co-tunneling line was observed as a result of two possible sets of transitions: 
within the pair $(4,2)$ or $(4,1)$ for positive, and within pair $(3,1)$ or 
$(3,2)$ for negative field orientation [cf. Fig.~\ref{fig:elevels}(b)]. In our 
case only the transitions between the ${\mathcal{C}}$-conjugated states, $(4,2)$ 
and $(3,1)$, are observed while the transitions between the 
${\mathcal{P}}$-conjugated states, $(4,1)$ and $(3,2)$, are absent. This 
observation substantiates the previous experimental report in \cite{Quay2007}, 
and is in nearly perfect agreement with the results of our many-body theory 
plotted in Fig.~\ref{fig:magnetic}(c). No Kondo-enhancement of the virtual 
transitions (4,1) and (3,2) occurs as a consequence of the symmetry constraints 
imposed onto the Keldysh action discussed in the following 
Sec.~\ref{sec:modeling}. These constraints reduce the allowed number of Kondo 
peaks expected in a perpendicular magnetic field with respect to earlier 
theoretical predictions (cf. Refs. \cite{Fang2008, Fang_Erratum}).

In magnetic fields parallel to the tube axis the satellite Kondo peaks are 
expected to move and split because, according to Figs.~\ref{fig:elevels}(b),(c), 
the single particle states 2 and 4 (1 and 3) are mutually tilted by the 
Aharonov-Bohm effect. Inspection of Figs.~\ref{fig:magnetic}(d) and 
\ref{fig:magnetic}(e) shows that the Kondo satellites now depend on the 
magnetic field, in qualitative agreement with our theoretical result displayed 
in Fig.~\ref{fig:magnetic}(f). Note that the following parameters of our model 
Hamiltonian are extracted from the experimental data: the ratio $\Delta/\kTK$ 
(see Fig.~\ref{fig:tempdep}), and the ratio $\DSO/\kTK$, which are obtained 
from the evolution of the splitted central Kondo peaks according to 
Eqs.~(\ref{eps1_eps_2_perp_low}) and (\ref{eps1_eps2_high_par}), respectively. 
For the parallel field case only, also the ratio $g_{\text{orb}} / 
g_{\text{s}}$ of orbital and spin $g$-factor has to be set. The parameters used 
to generate the theoretical curves in Figs.~\ref{fig:magnetic}(c) and 
\ref{fig:magnetic}(f) are given in Table~\ref{values} of 
Appendix~\ref{app:singleparticle}.

In order to better identify which transitions contribute to the evolution of the 
central peaks and of the satellites at finite magnetic fields, we compare the 
maxima of the $G(\vsd)$ traces [orange, blue and grey dots in 
Fig.~\ref{fig:comparison}] with the results of the many body theory and 
the energy differences of the underlying single particle levels [thick and thin 
lines  in Fig.~\ref{fig:comparison}, respectively]. Orange and blue dots 
correspond to transitions that, according to theory, are of the 
${\mathcal{T}}$- and ${\mathcal{C}}$-type, respectively. To the grey dots no 
clear assignment can be made~\footnote{For the theoretical calculations in 
Figs.~\ref{fig:comparison}(a) and \ref{fig:comparison}(c) we use the same 
independently determined parameters as in Fig.~\ref{fig:magnetic}. To fit the 
high field data in Fig.~\ref{fig:comparison}(b) a smaller orbital $g$-factor 
than at low field was used (see Table~\ref{values}).}. Noticeably, only 
transitions of the ${\mathcal{T}}$- and  ${\mathcal{C}}$-type are seen. A 
careful inspection of the second derivative of the $I(\vsd)$ confirms that the 
lines corresponding to transitions between ${\mathcal{P}}$-conjugated states are 
indeed absent in the experiment.  The absence of a splitting of the satellite 
peaks in the perpendicular field is very prominent in 
Fig.~\ref{fig:comparison}(a), where the position of the satellite peaks is 
essentially free of dispersion.  On the other hand, the satellite peaks do split 
in a parallel magnetic field, where the Aharonov-Bohm effect acts differently on 
the two pairs of ${\mathcal{C}}$-conjugated states as seen in 
Fig.~\ref{fig:comparison}(b). The high field behavior observed here is similar 
to that in \cite{Lan2012}, and confirms the absence of $\mathcal{P}$-transitions 
reported there. An additional discussion of the evolution of the conductance 
peaks can be found in Appendix~\ref{app:inelastic}.

Finally, we focus on the critical behavior of the central Kondo peaks in a 
magnetic field $\vec{B}$. A single Kondo peak is expected as long as the level
separation of the underlying single particle states does not exceed the 
threshold value $\Ecrit$. The threshold value \Ecrit\ (dotted red lines in 
Fig.~\ref{fig:comparison}(a), (c)) defines one critical magnetic field 
$B_{\text{c}}=0.6\,\kTK/g_{\text{s}}\muB$ for the perpendicular, and three 
critical fields $B_{\text{c1}}$ -- $B_{\text{c3}}$ for the parallel field 
direction. This characteristic difference arises from the additional crossing of 
the single particle levels 3 and 4 near $\kTK /g_{\text{s}}\muB\simeq 3.3\un{T}$ 
in Fig.~\ref{fig:elevels}(b), see also Ref. \cite{Jespersen2011}.
Results of our many-body calculations (thick lines in panels (a) and (c)) 
well match our experiment in perpendicular and in low  parallel magnetic fields. 
The non-linear dispersion of the positions of the central  Kondo peaks reflects 
the protection of the Kondo state against perturbations on energy scales below 
\kTK.

\section{Modeling and nonlinear transport theory}
\label{sec:modeling}
To account for the striking findings in magnetic field, we have developed a 
nonequilibrium field theory based on the slave boson Keldysh effective action 
formalism \cite{Altland_2010}. An \SUT\ formulation of this theory has been 
presented in Refs.~\cite{Smirnov_2013, Smirnov_2013a} and an \SUF\ formulation 
(including a broken \SUF) is presented in this work in 
Appendix~\ref{app:manyparticle}. The theory is based on the minimal model 
Hamiltonian $\hat{H}_{\text{CNT}}$ for a single longitudinal mode of a CNT 
quantum dot in magnetic field as given in Eq.~(\ref{Ham_general}) of the 
Appendix. Accounting for the four eigenstates $\left\lbrace |j \rangle, 
~j=1,2,3,4\right\rbrace $ of $\hat{H}_{\text{CNT}}$, the Coulomb interaction 
among them, and assuming a tunneling coupling which preserves CNT quantum 
numbers, the theory provides an approximate analytical expression for the four 
contributions $\nu_j(\varepsilon)$ to the tunneling density of states (TDOS)  
of the quantum dot. This expression is then used to evaluate the differential 
conductance as a function of the temperature, bias voltage and magnetic field 
over the whole energy range relevant for Kondo physics according to the 
Meir-Wingreen formula \cite{Wingreen_1994}, cf. Eq.~(\ref{MW}) of the Appendix. 

To explicit calculate  $\nu_j(\varepsilon)$, the interacting CNT Hamiltonian is 
first expressed in terms of slave bosons and fermions. The fermions in the tube 
and in the leads are then integrated out, leaving a still exact expression for 
$\nu_j(\varepsilon)$  in terms of the Keldysh effective action, see 
Eq.~(\ref{KFI_observ}) and (\ref{F_TDOS}), which only depends on the slave 
boson fields. In the limit of infinite charging energy, they represent 
fluctuations of the empty state of the dot. The bosonic fields cannot be 
integrated out exactly as the tunneling term of the Keldysh action, 
Eq.~(\ref{tun_act}), is nonlinear in them. Crucially though, this tunneling 
action is constructed in a such a way that for each CNT level $i$ there are two 
expansion points ($\gamma_i$ and $\delta_i$  in Eqs. (\ref{tun_matr}), 
(\ref{tun_matr_conj}), respectively).  Upon expanding the action  around the 
expansion points and retaining only quadratic terms, each $\nu_j$  is readily 
obtained by functional integration over the slave bosonic fields.  

The essence of the Kondo physics is the enhancement of certain virtual 
transitions when going towards low energies. In the perturbative regime the 
enhancement is only logarithmic. Thus one expects that all the transitions 
(independent of whether they are logarithmically enhanced or not) should be 
experimentally accessible by inelastic cotunneling spectroscopy  (see e.g. 
\cite{Jespersen2011}). In the low energy non perturbative regime the enhancement 
is much larger, as it yields resonances on the order of $e^2/h$. Thus, it is 
only in the latter regime that the lack of enhancement of some transitions 
clearly appears. A commonly used approach to determine which  transitions are 
enhanced  is to solve flow equations for the associated coupling constants in an 
effective Kondo model, see e.g. Refs. \cite{Paaske2006,Lim2006} for an 
application to CNTs. As outlined above, in the Keldysh approach one is not 
solving flow equations for the coupling constants. Rather, the evolution towards 
low energies is controlled by the expansion points $\gamma_i$ and $\delta_i$ in 
the effective Keldysh action. The expansion points $\gamma_i$ and 
$\delta_i$ are free parameters of the theory. Their value is fixed $ a\ 
posteriori$ by imposing the proper low energy behavior of the total TDOS  as 
known e.g. from Fermi liquid theory  \cite{expansion_point}. Additionally, they 
are fixed in such a way that  conjugation relations for the single particle 
spectrum are reflected in analogous conjugation relations among the  $\nu_j$.
Notice that for each $\nu_j$ the four complex quantities $E_i^j 
= \gamma_i^j\delta_i^j$ have to be determined.

For the \SUT\ Anderson model at zero magnetic field, time reversal symmetry 
requires $\nu_1=\nu_2$ ($j=1,2$ for the two spin degenerate levels). Hence, 
$E_i^j=E \; \forall \; i,j=1,2$. Then, the value of the complex quantity $E$ is 
{\it uniquely} fixed by constraints on the tunneling density of states and its 
derivative at zero temperature and at the Fermi energy known from Fermi liquid 
theory. A very good agreement of the theory with equilibrium numerical 
renormalization group results and with out of equilibrium real-time 
renormalization group predictions was demonstrated over the whole parameter 
regime \cite{Smirnov_2013}. To account for the effects of finite magnetic 
fields, a spin dependence of the expansion points must be included. In Ref. 
\cite{Smirnov_2013a} the choice was dictated by the observation that a) the 
flow towards energies requires virtual spin-flip processes (see e.g. 
\cite{Altland_2010}), and b) conjugation relations among the magnetic field 
splitted single particle levels should also be reflected at the level of the 
TDOS, i.e. $\nu_1(B)=\nu_2(-B)$. According to a) and b), the choice of the 
expansion points is such that each of the two contributions $\nu_1$ and $\nu_2$ 
effectively contains (through the related self-energy) only virtual spin-flip 
processes. This together with the Fermi liquid conditions on the TDOS at zero 
field uniquely determines the value of the expansion points. The merit of this 
choice has to be checked against other theories or experimental findings. As 
shown in Ref. \cite{Smirnov_2013a} (cf. Fig. 3 there), the theory well 
reproduces the experimentally observed evolution of the split Kondo peaks in a 
magnetic field seen in Ref.~\cite{Quay2007}. However, it does not 
quantitatively describe the tails of the Zeeman split peaks, as some inelastic 
cotunneling contributions are not retained in the second order expansion of the 
effective action. 

A similar reasoning is applied in this work. Due to the presence of orbital and 
spin degrees of freedom, the total TDOS is the sum of the four contributions 
$\nu_j(\varepsilon)$, one for each of the four levels.  For each $\nu_j$, the 
expansion points of the Keldysh effective action, i.e. the real and imaginary 
parts of the four $E^j_i$ are chosen according to the requirement 
that a) virtual processes that flip the quantum number are crucial for the low 
energy behavior \cite{Choi2005, Lim2006}, b) the TDOSs should be related by 
conjugation relations inherited from the single particle spectrum  reflecting 
$\mathcal{T}$-  and $\mathcal{P}$-conjugation in finite magnetic field:
\begin{equation}
\nu_{1}(\vec{B})=\nu_{2}(-\vec{B}),  \quad \nu_3(\vec{B})=\nu_4(-\vec{B}), \quad 
\nu_1(\Delta)=\nu_4(-\Delta), 
\end{equation}
where $\Delta(\vec{B})$ is the magnetic field dependent inelastic energy 
introduced in Sec.~\ref{sec:conjugation}. These requirements, and the $\SUF$ 
limit at zero field $\vec{B}$ and zero splitting $\Delta$, $uniquely \ fix$  
the expansion points at low energies. One crucial consequence of the combined 
quantum number flip and conjugations requirement, is that the TDOS $\nu_j$ 
effectively only contains (through its self-energy $\Sigma_j$) virtual 
transitions to the $\mathcal{T}$ and $\mathcal{C}$-conjugated partners of the 
level $j$, but not to the $\mathcal{P}$-conjugated ones. The explicit form for 
the $E_i^j$, $\nu_j$, and $\Sigma_j$ is given in Eqs. (\ref{expansionpoints}), 
(\ref{TDOS}), and (\ref{S_e2}) in the Appendix.

\section{Conclusion}
In conclusion, our work provides a  systematic experimental and theoretical 
investigation of the Kondo effect in carbon nanotubes in the presence of both 
spin-orbit coupling and valley mixing. The wide tunability of the carbon 
nanotube spectrum by magnetic fields allows to elucidate the role of symmetry 
and conjugation relations. Despite the symmetry breaking by spin-orbit 
interaction and valley mixing, the underlying operators still give rise 
to conjugation relations between certain states that have to be respected by 
the transport theory. The interplay of electron-electron interactions and 
conjugation relations lead  to the enhancement of selected many-body transitions 
only.  This explains the unexpected absence of several resonances in the 
nonequilibrium  Kondo transport spectrum in the non perturbative regime.

\section*{Acknowledgments}
The authors acknowledge financial support by the Deutsche 
Forschungsgemeinschaft via Emmy Noether project Hu~1808/1, SFB~631, SFB~689, 
and GRK~1570, the EU within SE2ND, and by the Studienstiftung des deutschen 
Volkes.

\appendix

\section{Single-particle energies and eigenstates in CNTs: Symmetries and 
conjugation operations of the effective CNT-Hamiltonian}
\label{app:singleparticle}

To understand the nonequilibrium many-particle physics of carbon nanotubes 
(CNTs), it is essential to first analyze the symmetry properties of the 
underlying single-particle Hamiltonian. Below the single-particle states of a 
quantum dot made of a carbon nanotube with the curvature induced spin-orbit 
interaction and valley mixing are presented.

The principal degrees of freedom characterizing the low energy states in a
carbon nanotube are, for given longitudinal mode, the longitudinal momentum 
$k$, the spin ($\sigma=\pm 1=\ \uparrow, \downarrow$), and the orbital 
pseudospin ($\tau=\pm 1=K,K'$) commonly referred to as {\em valley}. The valley 
labels $K,K'$ correspond to the clockwise and counterclockwise motion of the 
electrons around the CNT. Hence, to a given longitudinal mode a quadruplet of 
states in the composite spin and pseudospin space is associated.

Carbon nanotubes display several physical effects involving spin and valley 
degrees of freedom. Very prominent is the curvature induced spin-orbit 
interaction (SOI) \cite{Ando2000, Huertas-Hernando2006, Izumida2009, 
prb-delvalle:165427}. It breaks the four-fold spin and valley degeneracy and
splits the quartet of states into two Kramers doublets, separated in energy by
\DSO, with parallel and antiparallel alignment of spin and orbital magnetic 
moment. The SOI defines a preferred quantization axis for the spin (along the 
axis of the nanotube) and a certain composition in the valley space (pure 
valley eigenstates). Thus the natural eigenstate basis of an infinite CNT 
without the magnetic field is provided by the set $\{\vert K\uparrow\rangle, 
\vert K\downarrow\rangle,\vert K'\uparrow \rangle,\vert K'\downarrow\rangle\}$.

In finite CNTs the boundaries can induce a mixing between the two valleys: as
the reflection off the boundaries must reverse the axial momentum of the
particle, it can enforce a change of valley. The resulting hybridization of
different valley eigenstates introduces an energy difference \DKK\ between their
bonding ($\vert b\rangle = (-\vert K\rangle + \vert K'\rangle)/\sqrt{2}$)
and antibonding ($\vert a\rangle = (\vert K\rangle + \vert K'\rangle) /
\sqrt{2}$) combinations \cite{Jespersen2011}. The valley-mixing term acts 
therefore against \DSO, which favors pure valley eigenstates. When \DKK\ is 
included, a convenient basis in the valley space becomes that of the mixed 
valley eigenstates $\vert b\rangle$ and $\vert a\rangle$.

In a finite magnetic field $\vec{B}$ the Zeeman effect splits the energies of 
the spin parallel and antiparallel to the magnetic field by  $g_{\text{s}}\muB 
B$, with $B=|\vec{B}|$. Thus it favors the direction of the field as the spin 
quantization axis. If the field is parallel to the CNT axis, the Zeeman effect 
cooperates with the SOI in that that the spin still remains a good quantum 
number. For any other field direction the Zeeman effect and \DSO\ compete 
against each other, the former trying to align the spins with the field, the 
latter with the CNT axis. 

If the field has a non-vanishing parallel component $B_\parallel$, the 
Aharonov-Bohm effect is induced by the cylindrical topology of the CNT. This 
alters the energies of the two valley eigenstates, raising the energy of one 
and lowering the energy of the other. The energy gap between the two valleys is 
$2g_{\text{orb}}(B)\muB \Bpar$, with $g_{\text{orb}}$ typically larger than 
$g_{\text{s}}$.

The minimal Hamiltonian of a CNT quadruplet in the presence of a magnetic field 
of strength $B$, applied at an angle $\varphi$ to the CNT axis, and written in 
the basis $\{\vert a\uparrow\rangle,\vert b\uparrow\rangle, \vert 
a\downarrow\rangle,\vert b\downarrow\rangle\}$ is then \cite{Jespersen2011}
\begin{equation}
\begin{split}
\hat{H}_{\text{CNT}}= & \varepsilon_d
\,\hat{I}_{\sigma}\otimes\hat{I}_{\tau}
+ \frac{\DKK}{2}\hat{I}_{\sigma}\otimes\hat{\tau}_z+\frac{\DSO}{2} 
\hat{\sigma}_z \otimes\hat{\tau}_x \\
+ & \frac{1}{2}g_{\text{s}}\mu_{\text{B}}|\vec{B}| 
\left(\cos\varphi\,\hat{\sigma}_z 
+ \sin\varphi\,\hat{\sigma}_x\right) \otimes\hat{I}_{\tau} \\
+ & g_{\text{orb}}\mu_{\text{B}}|\vec{B}| \cos\varphi\; 
\hat{I}_{\sigma}\otimes\hat{\tau}_x
.
\end{split}
\label{Ham_general}
\end{equation}
The operators $\hat{\tau}_i$ and $\hat{\sigma}_i$, $i=x,y,z$, act in the valley
and spin spaces, respectively. The states $|a\rangle$ and $|b\rangle$
are the eigenstates of $\hat{\tau}_z$ corresponding to the eigenvalues $+1$
and $-1$, respectively, while $|\uparrow\rangle$ and $|\downarrow\rangle$ are 
the eigenstates of $\hat{\sigma}_z$ corresponding to its eigenvalues $+1$ and 
$-1$. The spin and orbital magnetic moments are given by 
$\frac{1}{2}g_{\text{s}}\mu_B$ and $g_{\text{orb}}\mu_B$ respectively, and 
$\varepsilon_d$ is a reference energy for the considered longitudinal mode.

This Hamiltonian has eigenstates $\{\vert 1 \rangle,\vert 2 \rangle, \vert 3 
\rangle,\vert 4 \rangle\}$ of energies $\varepsilon_i = \varepsilon_i 
(\vec{B})$, $i=1,2,3,4$. In the following we shall explicitly introduce three 
operators $\hat{\mathcal{T}}$, $\hat{\mathcal{P}}$, $\hat{\mathcal{C}}$ which 
enable to conjugate states of the quadruplet pairwise. We shall focus first on 
the case of zero field and then on the two special physical cases relevant for 
our experiments, {\it i.e.}, when the magnetic field is perpendicular to the 
carbon nanotube axis and when it is parallel to it. A summary of the 
conjugation considerations is given at the end of 
Appendix~\ref{app:singleparticle}.

\subsection{Zero field}
The analysis of the spectrum of the Hamiltonian Eq.~(\ref{Ham_general}) at zero 
magnetic field, $\hat{H}^{(0)}_{\text{CNT}}$, is crucial for the understanding 
of the implications of valley mixing and spin-orbit coupling on the CNT 
spectrum. It reads
\begin{equation}
\hat{H}^{(0)}_{\text{CNT}}= \varepsilon_d
\,\hat{I}_{\sigma}\otimes\hat{I}_{\tau}
+ \frac{\DKK}{2}\hat{I}_{\sigma}\otimes\hat{\tau}_z
+ \frac{\DSO}{2} \hat{\sigma}_z \otimes\hat{\tau}_x.
\label{Ham_zero_B}
\end{equation}
The eigenstates $\{\vert 1\rangle,\vert2\rangle,\vert 3\rangle,\vert 4\rangle\}$ 
can be easily expressed in terms of the bonding/antibonding states according to
\begin{equation}
\begin{bmatrix}
|1\rangle\\|4\rangle\\|2\rangle\\|3\rangle
\end{bmatrix}=
\begin{bmatrix}
\cos\bigl(\frac{\theta}{2}\bigl)&\sin\bigl(\frac{\theta}{2}\bigl)&0&0\\
-\sin\bigl(\frac{\theta}{2}\bigl)&\cos\bigl(\frac{\theta}{2}\bigl)&0&0\\
0&0&\cos\bigl(\frac{\theta}{2}\bigl)&-\sin\bigl(\frac{\theta}{2}\bigl)\\
0&0&\sin\bigl(\frac{\theta}{2}\bigl)&\cos\bigl(\frac{\theta}{2}\bigl)\\
\end{bmatrix}
\begin{bmatrix}
|a\uparrow\rangle\\|b\uparrow\rangle\\|a\downarrow\rangle\\|b\downarrow\rangle
\end{bmatrix},
\label{Eig_bas_rel}
\end{equation}
where $\tan(\theta)=\DSO/\DKK$. Due to spin conservation 
($[\hat{H}^{(0)}_\text{CNT},\hat{\sigma}_z] =0$), the unitary matrix connecting 
the two basis sets is block diagonal and only mixes the valley degrees of 
freedom. Diagonalization of $\hat{H}^{(0)}_\text{CNT}$ yields $\varepsilon_1 = 
\varepsilon_2$, $\varepsilon_3 = \varepsilon_4$, and $\varepsilon_1 = 
\varepsilon_3+\Delta$, where $\Delta=\sqrt{\DKK^2+\DSO^2}$.

\subsubsection{Time-reversal symmetry}
Let us now investigate the action of the anti-unitary time-reversal operator 
$\hat{\mathcal{T}}$,
\begin{equation}
\hat{\mathcal{T}}=-i\hat{\sigma}_y\otimes\hat{\tau}_z \kappa,
\end{equation}
where $\kappa$ stands for complex conjugation. The operator $\mathcal{T}$ 
commutes with $\hat{H}^{(0)}_\text{CNT}$:
\begin{equation}
\hat{\mathcal{T}}\hat{H}^{(0)}_\text{CNT}\hat{\mathcal{T}}^{-1}=
\hat{H}^{(0)}_\text{CNT},
\; \mathrm{or} \; [\hat{\mathcal{T}},\hat{H}^{(0)}_\text{CNT}]=0\;,
\label{TH0}
\end{equation} 
i.e. the CNT Hamiltonian has time-reversal (TR) symmetry which implies 
doublets of energy degenerate states (Kramers pairs). The 
$\hat{\mathcal{T}}$-conjugated pairs of states are easily identified to be 
$\{\vert 1\rangle,\vert2\rangle\}\equiv(1,2)_T$ and $\{\vert 
3\rangle,\vert4\rangle\}\equiv(3,4)_T$ due to
\begin{equation}
\begin{split}
\hat{\mathcal{T}}|1\rangle=\kappa |2\rangle,\qquad 
\hat{\mathcal{T}}|2\rangle=\kappa |1\rangle,\\
\hat{\mathcal{T}}|3\rangle=\kappa |4\rangle,\qquad
\hat{\mathcal{T}}|4\rangle=\kappa |3\rangle.
\label{TRSoperation}
\end{split}
\end{equation}
Notice that in agreement with the results form the diagonalization of 
$\hat{H}^{(0)}_\text{CNT}$ in Eq.~(\ref{Ham_zero_B}), Eq.~(\ref{TRSoperation}) 
implies $\varepsilon_1=\varepsilon_2$, $\varepsilon_3=\varepsilon_4$. We also 
identify the level splitting as $\Delta=\varepsilon_1-\varepsilon_3$.

\subsubsection{Particle-hole conjugation}
Let us now further proceed by introducing the anti-unitary operator 
$\hat{\mathcal{P}}$ associated to particle-hole conjugation within a given 
longitudinal mode:
\begin{equation}
\hat{\mathcal{P}}=\hat{\sigma}_z\otimes(-i\hat{\tau}_y)\kappa.
\end{equation}
This operator is constructed such that
\begin{multline}
\hat{\mathcal{P}}(\hat{H}^{(0)}_\text{CNT}-\varepsilon_d
\hat{I}_{\sigma}\otimes\hat{I}_{\tau})\hat{\mathcal{P}}^{-1}=-(\hat{H}^{(0)}
_\text{CNT}-\varepsilon_d
\hat{I}_{\sigma}\otimes\hat{I}_{\tau}), \\ 
\mathrm{or}\; 
\{\hat{\mathcal{P}},(\hat{H}^{(0)}_\text{CNT}-\varepsilon_d
\hat{I}_{\sigma}\otimes\hat{I}_{\tau})\}=0,
\label{PH0}
\end{multline}
i.e., the operators $\hat{\mathcal{P}}$ and 
($\hat{H}^{(0)}_\text{CNT}-\varepsilon_d
\hat{I}_{\sigma}\otimes\hat{I}_{\tau}$) \textit{anticommute}. The operator 
$\hat{\mathcal{P}}$ exchanges a state with an energy $\varepsilon$ above a 
certain reference energy $\varepsilon_d$ with the $\hat{\mathcal{P}}$-conjugated 
state with the energy $-\varepsilon$ below the reference energy.
In other words, the eigenenergies of the Hamiltonian (\ref{Ham_zero_B}) are 
exchanged under this transformation. The corresponding particle-hole conjugated 
pairs are $\{\vert 1\rangle,\vert4\rangle\}\equiv(1,4)_P$ and
$\{\vert 2\rangle,\vert3\rangle\}\equiv(2,3)_P$ as it follows from
\begin{equation}
\begin{split}
\hat{\mathcal{P}}\vert 1\rangle=\kappa \vert 4\rangle,\qquad
\hat{\mathcal{P}}\vert 2\rangle=- \kappa \vert 3\rangle,\\
\hat{\mathcal{P}}\vert 3\rangle=- \kappa \vert 2\rangle,\qquad
\hat{\mathcal{P}}\vert 4\rangle=\kappa \vert 1\rangle.
\label{P-transformation}
\end{split}
\end{equation}
It follows that $\varepsilon_1-\varepsilon_d
=-(\varepsilon_4-\varepsilon_d),\; \varepsilon_2-\varepsilon_d
=-(\varepsilon_3-\varepsilon_d)$. Moreover, combined with TR symmetry, this 
also implies $\varepsilon_1=\varepsilon_2=\varepsilon_d
+{\Delta}/{2}$, $\varepsilon_3=\varepsilon_4=\varepsilon_d
-{\Delta}/{2}$ and hence
\begin{equation}
\varepsilon_1(\Delta)=\varepsilon_4(-\Delta).
\end{equation}

\subsubsection{Chiral conjugation}
Chiral (C) conjugation is defined as a combination of $\hat{\mathcal{T}}$ and 
$\hat{\mathcal{P}}$ and given by the unitary operator 
\begin{equation}
\hat{\mathcal{C}}=\hat{\mathcal{P}}\hat{\mathcal{T}}^{-1}=\hat{\sigma}
_x\otimes\hat{\tau}_x.
\end{equation}
This implies that chiral conjugation holds if the $\hat{\mathcal{T}}$- and 
$\hat{\mathcal{P}}$-operations do. The corresponding conditions for chiral 
conjugation read:
\begin{multline}
\hat{\mathcal{C}}(\hat{H}^{(0)}_\text{CNT}-\varepsilon_d
\hat{I}_{\sigma}\otimes\hat{I}_{\tau})\hat{\mathcal{C}}^{-1}=-(\hat{H}^{(0)}
_\text{CNT}-\varepsilon_d
\hat{I}_{\sigma}\otimes\hat{I}_{\tau}), \; \mathrm{or} \\
\{\hat{\mathcal{C}},(\hat{H}^{(0)}_\text{CNT}-\varepsilon_d
\hat{I}_{\sigma}\otimes\hat{I}_{\tau})\}=0,
\label{CH0}
\end{multline}
i.e., the operators $\hat{\mathcal{C}}$ and 
($\hat{H}^{(0)}_\text{CNT}-\varepsilon_d
\hat{I}_{\sigma}\otimes\hat{I}_{\tau}$) also \textit{anticommute}.
The chiral pairs are $\{\vert 1\rangle,\vert3\rangle\}\equiv(1,3)_C$ and
$\{\vert 2\rangle,\vert4\rangle\}\equiv(2,4)_C$, as it follows from
\begin{equation}
\begin{split}
\hat{\mathcal{C}}\vert 1\rangle=\vert 3\rangle,\qquad
\hat{\mathcal{C}}\vert 2\rangle=\vert 4\rangle,\\
\hat{\mathcal{C}}\vert 3\rangle=\vert 1\rangle,\qquad
\hat{\mathcal{C}}\vert 4\rangle=\vert 2\rangle.
\end{split}
\end{equation}
It then holds $\varepsilon_1-\varepsilon_d
=-(\varepsilon_3-\varepsilon_d)$ and $\varepsilon_2-\varepsilon_d
=-(\varepsilon_4-\varepsilon_d)$.

The behavior of the eigenstates $\{\vert 1\rangle,\vert2\rangle, \vert 
3\rangle,\vert 4\rangle\}$ of $\hat{H}^{(0)}_\text{CNT}$ under the action 
of the operators $\hat{\mathcal{T}}, \hat{\mathcal{P}}, \hat{\mathcal{C}}$ is 
summarized in Fig.~\ref{fig:symmetries}(a). In the following we discuss how an 
external magnetic field affects these properties.

\subsection{Perpendicular magnetic field}
Let us start with the case of the perpendicular orientation. The Hamiltonian in
this case has the following form:
\begin{equation}
\hat{H}_\text{CNT}=\hat{H}^{(0)}_\text{CNT}+\hat{H}_\perp(\Bperp)=\hat{H}^{(0)}
_\text{CNT}+
\frac{1}{2}g_\text{s}\mu_\text{B}B_\perp\hat{\sigma}_x\otimes\hat{I}_{\tau}.
\label{Ham_perp}
\end{equation}
We now need to study the action of $\hat{\mathcal{T}}, \hat{\mathcal{P}}$, and 
$\hat{\mathcal{C}}$ on the magnetic field-dependent part of 
$\hat{H}_\text{CNT}$, i.e., 
$\hat{H}_\perp(\Bperp)=\frac{1}{2}g_\text{s}\mu_\text{B}B_\perp\hat{\sigma}
_x\otimes\hat{I}_{\tau}$.

\subsubsection{Conjugation under time-reversal}
We obtain:
\begin{multline}
\hat{\mathcal{T}}\,\hat{H}_\perp(\Bperp)\,\hat{\mathcal{T}}^{-1}=-\hat{H}
_\perp(\Bperp)=\hat{H}_\perp(-B_\perp),\; \mathrm{or} \\
\{\hat{\mathcal{T}},\hat{H}_\perp(\Bperp)\}=0\;.
\label{HamSymmTRS}
\end{multline}
Comparison with Eq.~(\ref{TH0}) lets us recognize that TR symmetry is now 
broken and hence the degeneracy within the Kramers pairs $(1,2)_T$ and 
$(3,4)_T$ is lifted. The last equality in the first half of 
Eq.~(\ref{HamSymmTRS}) implies that 
$\hat{\mathcal{T}}\vert 1,B_\perp\rangle=\kappa \vert 2,-B_\perp\rangle,~~
\hat{\mathcal{T}}\vert 3,B_\perp\rangle=\kappa \vert 4,-B_\perp\rangle$, if the 
eigenstates $\{\vert i,B_\perp\rangle,~i=1,2,3,4\}$ of $\hat{H}_\text{CNT}$ are 
taken from Eq.~(\ref{Ham_perp}). Correspondingly, the eigenenergies now obey
\begin{equation}
\varepsilon_1(B_\perp)=\varepsilon_2(-B_\perp),\qquad
\varepsilon_3(B_\perp)=\varepsilon_4(-B_\perp).
\label{Energy_perp}
\end{equation}

\subsubsection{Particle-hole conjugation}
In the magnetic field we find that
\begin{multline}
\hat{\mathcal{P}}\,\hat{H}_\perp(\Bperp)\,\hat{\mathcal{P}}^{-1}=-\hat{H}
_\perp(\Bperp)=\hat{H}_\perp(-B_\perp),\; \mathrm{or}\\
\{\hat{\mathcal{P}},\hat{H}_\perp(\Bperp)\}=0\;,
\label{HamSymmPHS}
\end{multline}
which implies
\begin{multline}
\hat{\mathcal{P}}(\hat{H}_\text{CNT}-\varepsilon_d
\hat{I}_\sigma\otimes\hat{I}_\tau)\hat{\mathcal{P}}^{-1}=-(\hat{H}_\text{CNT}
-\varepsilon_d
\hat{I}_\sigma\otimes\hat{I}_\tau),\; \mathrm{or} \\
\{\hat{\mathcal{P}},(\hat{H}_\text{CNT}-\varepsilon_d
\hat{I}_\sigma\otimes\hat{I}_\tau)\}=0\;.
\label{PH_sym_rel_perp}
\end{multline}
If we recall Eq.~(\ref{PH0}) we arrive at the important conclusion that the 
particle-hole (PH) conjugation {\em remains intact} in a perpendicular 
magnetic field. Thus Eq.~(\ref{P-transformation}) still holds and 
\begin{equation}
\begin{split}
\varepsilon_1(B_\perp)-\varepsilon_d
=-(\varepsilon_4(B_\perp)-\varepsilon_d),\\
\varepsilon_2(B_\perp)-\varepsilon_d
=-(\varepsilon_3(B_\perp)-\varepsilon_d).
\end{split}
\label{perpEnergyPHS}
\end{equation}

\subsubsection{Chiral conjugation}
Finally,
\begin{multline}
\hat{\mathcal{C}}\,\hat{H}_\perp(\Bperp)\,\hat{\mathcal{C}}^{-1}=
\hat{H}_\perp(\Bperp)=-\hat{H}_\perp(-B_\perp),\; \mathrm{or} \\
[\hat{\mathcal{C}},\hat{H}_\perp(\Bperp)]=0\;.
\label{HamSymmCS}
\end{multline}
Taking into account Eq.~(\ref{CH0}) we see that $\hat{\mathcal{C}}$ does no 
longer anticommute with $\hat{H}_\text{CNT}$, and that
\begin{equation}
\begin{split}
\varepsilon_1(B_\perp)-\varepsilon_d
=-(\varepsilon_3(-B_\perp)-\varepsilon_d),\\
\varepsilon_2(B_\perp)-\varepsilon_d
=-(\varepsilon_4(-B_\perp)-\varepsilon_d).
\end{split}
\label{perpEnergyCS}
\end{equation}

While the above conjugations (\ref{Energy_perp}), (\ref{perpEnergyPHS}), 
(\ref{perpEnergyCS}) are general because they are dictated by the (anti-) 
commutation relations (\ref{HamSymmTRS}), (\ref{HamSymmPHS}), 
(\ref{HamSymmCS}), 
they can be directly verified upon diagonalization of (\ref{Ham_perp}). 
The eigenstates $\{\vert i,B_\perp\rangle,~i=1,2,3,4\}$ are expressed in terms 
of the bonding/antibonding states as
\begin{equation}
\begin{split}
&
\begin{bmatrix}
|1,B_\perp\rangle\\|4,B_\perp\rangle\\|2,B_\perp\rangle\\|3,B_\perp\rangle
\end{bmatrix}=
A\;
\begin{bmatrix}
|a\uparrow\rangle+|a\downarrow\rangle\\|b\uparrow\rangle-|b\downarrow\rangle\\
|a\uparrow\rangle-|a\downarrow\rangle\\|b\uparrow\rangle+|b\downarrow\rangle
\end{bmatrix}, \quad\mathrm{where}\\[2mm]
&A =
\begin{bmatrix}
\cos\bigl({\theta^+}/{2}\bigl)&\sin\bigl({\theta^+}/{2}\bigl)&0&0\\
-\sin\bigl({\theta^+}/{2}\bigl)&\cos\bigl({\theta^+}/{2}\bigl)&0&0\\
0&0&\cos\bigl({\theta^-}/{2}\bigl)&\sin\bigl({\theta^-}/{2}\bigl)\\
0&0&-\sin\bigl({\theta^-}/{2}\bigl)&\cos\bigl({\theta^-}/{2}\bigl)\\
\end{bmatrix},
\end{split}
\label{Eig_bas_rel_perp}
\end{equation}
and
\begin{equation}
\tan(\theta^\pm)=\frac{\DSO}{\DKK\pm
g_\text{s}\mu_\text{B}B_\perp}.
\label{Omega}
\end{equation}
Equation~(\ref{Eig_bas_rel_perp}) represents rotations by angles $\theta^\pm/2$ 
in two independent planes, $\theta^\pm$-planes, involving the two particle-hole 
pairs $(1,4)_P$ and $(2,3)_P$, respectively. The corresponding eigenenergies 
are 
\begin{equation}
\boxed{
\begin{split}
&\varepsilon_{1,4}(B_\perp)= \varepsilon_d \pm 
\frac{1}{2}\Delta_\perp(\Bperp),\\
&\varepsilon_{2,3}(B_\perp)= \varepsilon_d \pm \frac{1}{2}\Delta_\perp(-\Bperp),
\end{split}}
\label{Eig_en_perp}
\end{equation}
with
\begin{equation} 
\Delta_\perp(B_\perp)=\sqrt{\DSO^2+(\DKK+g_\text{s}\mu_\text{B}B_\perp)^2}.
\label{Delta_perp}
\end{equation}
Notice that Eq.~(\ref{Eig_en_perp}) is still PH-symmetric. PH conjugation and 
the time-reversal equation (\ref{Energy_perp}) imply 
\begin{equation}
\begin{split}
\varepsilon_1\left[ \Delta_\perp(B_\perp)\right] &=\varepsilon_4\left[ 
-\Delta_\perp(B_\perp)\right] ,\\
\varepsilon_2\left[ \Delta_\perp(-B_\perp)\right] &=\varepsilon_3\left[ 
-\Delta_\perp(-B_\perp)\right].
\end{split}
\label{Energies_perp}
\end{equation}

\begin{figure}
\includegraphics[width=\columnwidth]{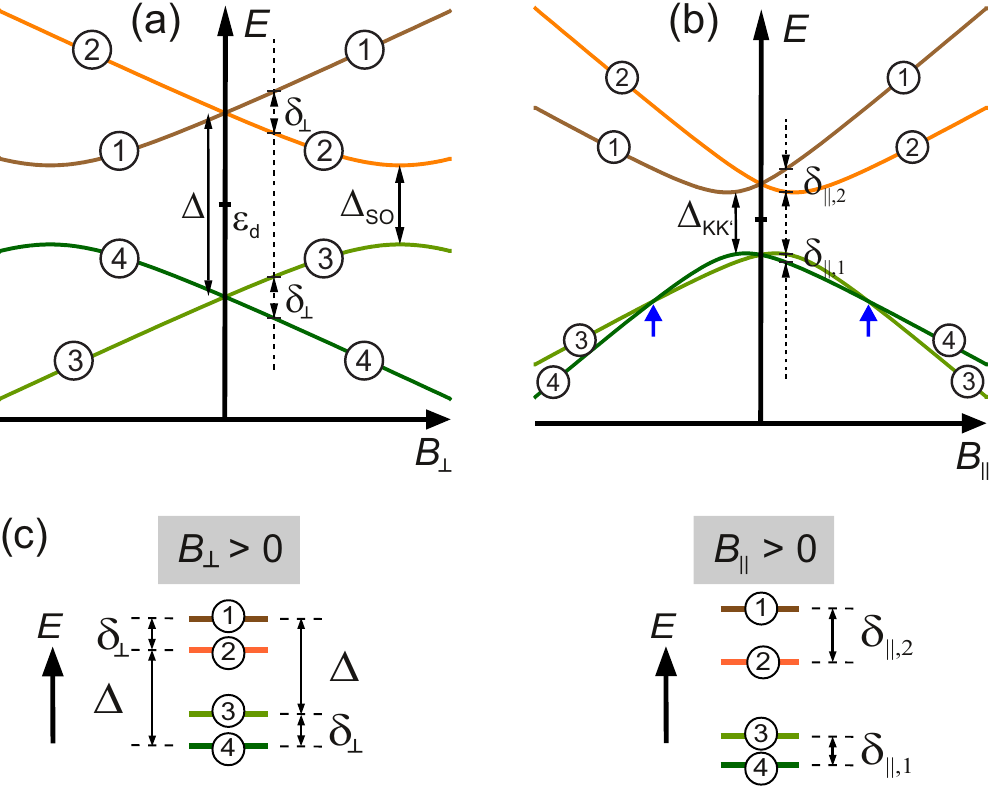}
\caption{\label{figure_e_level}(Color online) Sketch of the evolution of the 
energy levels in a magnetic field in the case of valley mixing $\DKK$ larger 
than the spin-orbit coupling $\DSO$. 
(a,b) Single particle spectrum in a magnetic field (a) 
perpendicular and (b) parallel to the CNT axis. The energy eigenstates are 
labeled 1-4. At $B=0$ there are two degenerate Kramers pairs separated by 
$\Delta$. (a) In perpendicular field the energy difference within each chiral 
pair, i.e.,  $(1,3)$ and $(2,4)$, remains essentially independent of \Bperp, 
provided that $\Bperp \lesssim \DKK/g_{\text{s}}\muB$. The level splitting 
within each Kramers doublet is $\delta_\perp=g'_s\mu_B\Bperp$.  An avoided 
crossing of the particle-hole pair (2,3) occurs when $\Bperp \simeq \DKK / 
g_{\text{s}}\muB$. (b) For $\Bpar$ the Aharonov-Bohm effect induces a level 
crossing of the pair $(3,4)$, as indicated by the blue arrows. (c) 
Visualization of the level separations from (a) and (b) for values of $\Bperp$ 
and $\Bpar$ corresponding to the dashed lines.
 }
\label{fig:StateEvol}
\end{figure}

The evolution of the four states in perpendicular field is shown in 
Fig.~\ref{fig:StateEvol}(a) together with the energy scale $\varepsilon_d$. The 
meaning of $\Delta_\perp(\Bperp)$ and $\Delta_\perp(-\Bperp)$ is visualized in 
Fig.~\ref{fig:elevels}(a).

One can easily obtain from Eq.~(\ref{Eig_en_perp}) the low and high field 
asymptotics. The difference $\varepsilon_1(B_\perp)-\varepsilon_2(B_\perp)$ is 
particularly relevant for our experiments, because at small fields one can 
extract the effective $g$-factor 
\begin{equation}
\begin{split}
\delta_\perp&=\varepsilon_1(B_\perp)-\varepsilon_2(B_\perp)\simeq 
g'_\text{s}\mu_\text{B}B_\perp,\\
g'_\text{s}&\equiv\frac{g_\text{s}}{\sqrt{1+\bigl(\Delta_\text{SO}/\Delta_\text{
KK'}\bigl)^2}},
\end{split}
\label{eps1_eps_2_perp_low}
\end{equation}
where $g'_\text{s}$ is the experimentally measured effective $g$-factor. At 
large fields, $g_\text{s}\mu_\text{B}B_\perp\gg\sqrt{\DKK^2+\DSO^2}$, 
\begin{equation}
\varepsilon_1(B_\perp)-\varepsilon_2(B_\perp)\simeq\Delta_\text{KK'}
\label{eps1_eps_2_perp_high}
\end{equation}
becomes field independent, providing a direct way to measure 
$\Delta_\text{KK'}$.

\subsection{Parallel magnetic field}
In the case of the magnetic field oriented along the axis of the carbon nanotube
the Hamiltonian takes the form
\begin{multline}
\hat{H}_\text{CNT}=
\hat{H}^{(0)}_\text{CNT}+\hat{H}_\parallel(\Bpar)=
\hat{H}^{(0)}_\text{CNT}+
g_\text{orb}\mu_\text{B}B_\parallel\hat{I}_{\sigma}\otimes\hat{\tau}_x+\\
\frac{1}{2}g_\text{s}\mu_\text{B}B_\parallel\hat{\sigma}_z\otimes\hat{I}_{\tau},
\label{Ham_par}
\end{multline}
where $B_\parallel$ is the magnitude of the parallel magnetic field and
$g_\text{orb}$ is the orbital $g$-factor.

Let us in a similar way address the action of $\hat{\mathcal{T}}, 
\hat{\mathcal{P}}$, and $\hat{\mathcal{C}}$ on the magnetic field dependent 
part of $\hat{H}_\text{CNT}$, i.e., $\hat{H}_\parallel(\Bpar) = g_\text{orb} 
\mu_\text{B} B_\parallel\hat{I}_{\sigma} \otimes\hat{\tau}_x + 
\frac{1}{2}g_\text{s}\mu_\text{B} B_\parallel \hat{\sigma}_z\otimes 
\hat{I}_{\tau}$.

\subsubsection{Conjugation under time-reversal}
In analogy to Eqs.~(\ref{HamSymmTRS}), (\ref{Energy_perp}) we find
\begin{multline}
\hat{\mathcal{T}}\,\hat{H}_\parallel(\Bpar)\,\hat{\mathcal{T}}^{-1}=
-\hat{H}_\parallel(\Bpar)=\hat{H}_\parallel(-\Bpar),\; \mathrm{or} \\
\{\hat{\mathcal{T}},\hat{H}_\parallel(\Bpar)\}=0\;,
\end{multline}
while the eigenstates $\{\vert i,B_\parallel\rangle,~i=1,2,3,4\}$ now obey
$\hat{\mathcal{T}}\vert 1,B_\parallel\rangle=
\kappa\vert 2,-B_\parallel\rangle,~\hat{\mathcal{T}}\vert 3,B_\parallel\rangle=
\kappa\vert 4,-B_\parallel\rangle$. 
Correspondingly, the eigenenergies are related through
\begin{equation}
\varepsilon_1(B_\parallel)=\varepsilon_2(-B_\parallel),\qquad
\varepsilon_3(B_\parallel)=\varepsilon_4(-B_\parallel).
\label{Bparallel_symm}
\end{equation}

\subsubsection{Particle-hole conjugation}
If we now look at the action of $\hat{\mathcal{P}}$ we observe
\begin{multline}
\hat{\mathcal{P}}\,\hat{H}_\parallel(\Bpar)\,\hat{\mathcal{P}}^{-1}=
-\hat{H}_\parallel(\Bpar)=\hat{H}_\parallel(-\Bpar),\; \mathrm{or} \\
\{\hat{\mathcal{P}},\hat{H}_\parallel(\Bpar)\}=0\;,
\label{Particle-symm}
\end{multline}
which implies
\begin{equation}
\begin{split}
\hat{\mathcal{P}}(\hat{H}_\text{CNT}-\varepsilon_d
\hat{I}_\sigma\otimes\hat{I}_\tau-\frac{1}{2}g_\text{s}\muB \Bpar 
\hat{\sigma}_z\otimes\hat{I}_\tau)\hat{\mathcal{P}}^{-1} \\
=-(\hat{H}_\text{CNT}
-\varepsilon_d
\hat{I}_\sigma\otimes\hat{I}_\tau-\frac{1}{2}g_\text{s}\muB \Bpar 
\hat{\sigma}_z\otimes\hat{I}_\tau),\quad \\[2mm] \mathrm{or}\quad 
\{\hat{\mathcal{P}},(\hat{H}_\text{CNT}-\varepsilon_d
\hat{I}_\sigma\otimes\hat{I}_\tau-\frac{1}{2}g_\text{s}\muB \Bpar 
\hat{\sigma}_z\otimes\hat{I}_\tau)\}&=0\;.
\label{PH_sym_rel_par}
\end{split}
\end{equation}
We see that PH symmetry is still obeyed if we measure the energies with respect 
to the reference energy $\varepsilon(\Bpar) = \varepsilon_d + \frac{1}{2} 
g_\text{s}\mu_\text{B}B_\parallel$, which now depends on $\Bpar$ (cf. 
Eq.~(\ref{PH0})). Furthermore Eq.~(\ref{P-transformation}) holds also in 
parallel magnetic field and hence
\begin{equation}
\begin{split}
\varepsilon_1(\Bpar)-\varepsilon(\Bpar)=
-\varepsilon_4(\Bpar)+\varepsilon(\Bpar),\\
\varepsilon_2(\Bpar)-\varepsilon(-\Bpar)=
-(\varepsilon_3(\Bpar)-\varepsilon(-\Bpar)) \;.
\end{split}
\end{equation}
With $\Delta(\Bpar)=2(\varepsilon_1(\Bpar)-\varepsilon(\Bpar))$ one finds:
\begin{equation}
\begin{split}
\varepsilon_1\left[ \varepsilon
(B_\parallel),\Delta(B_\parallel)\right] &=
\varepsilon_4\left[ \varepsilon
(B_\parallel),-\Delta(B_\parallel)\right] ,\\
\varepsilon_2\left[ \varepsilon
(-B_\parallel),\Delta(-B_\parallel)\right] &=
\varepsilon_3\left[ \varepsilon
(-B_\parallel),-\Delta(-B_\parallel)\right] .
\end{split}
\end{equation}

\subsubsection{Chiral conjugation}
Regarding chiral conjugation
\begin{equation}
\begin{split}
\hat{\mathcal{C}}\,\hat{H}_\parallel(\Bpar)\,\hat{\mathcal{C}}^{-1}=
-\hat{H}_\parallel(\Bpar)=\hat{H}_\parallel(-\Bpar),\; \mathrm{or} \\
[\hat{\mathcal{C}},\hat{H}_\parallel(\Bpar)]=0\;,
\end{split}
\end{equation}
we see that $\hat{\mathcal{C}}$ and $\hat{H}_\parallel(\Bpar)$ commute. Taking 
into account Eq.~(\ref{CH0}) it follows that $\hat{\mathcal{C}}$ again does not 
commute with $\hat{H}_{\text{CNT}}$. The symmetry relations between energies of 
the $\hat{\cal C}$-conjugated states are analogous to those in 
Eq.~(\ref{perpEnergyCS}). Along similar lines as for the 
$\hat{\mathcal{P}}$-operation one then finds the relation $\varepsilon_1\left[ 
\varepsilon(\Bpar),\Delta_\parallel(\Bpar)\right] =\varepsilon_3\left[ 
\varepsilon(-\Bpar),-\Delta_\parallel(-\Bpar)\right] $. 

The eigenstates $\{\vert i,B_\parallel\rangle,~i=1,2,3,4\}$ of the Hamiltonian 
(\ref{Ham_par}) are easily obtained by taking into account that in parallel 
field the spin states $\{\vert \uparrow\rangle,\vert \downarrow\rangle\}$ are 
still eigenstates of the total Hamiltonian (\ref{Ham_par}). We find 
\begin{equation}
\begin{split}
& \begin{bmatrix}
|1,B_\parallel\rangle\\|4,B_\parallel\rangle\\|2,B_\parallel\rangle\\|3,
B_\parallel\rangle
\end{bmatrix}= A
\begin{bmatrix}
|a\uparrow\rangle\\|b\uparrow\rangle\\|a\downarrow\rangle\\|b\downarrow\rangle
\end{bmatrix}, \quad\mathrm{where}\\[2mm]
& A = \begin{bmatrix}
\cos\bigl(\frac{\Theta^+}{2}\bigl)&\sin\bigl(\frac{\Theta^+}{2}\bigl)&0&0\\
-\sin\bigl(\frac{\Theta^+}{2}\bigl)&\cos\bigl(\frac{\Theta^+}{2}\bigl)&0&0\\
0&0&\cos\bigl(\frac{\Theta^-}{2}\bigl)&-\sin\bigl(\frac{\Theta^-}{2}\bigl)\\
0&0&\sin\bigl(\frac{\Theta^-}{2}\bigl)&\cos\bigl(\frac{\Theta^-}{2}\bigl)\\
\end{bmatrix},
\end{split}
\label{Eig_bas_rel_par}
\end{equation}
and
\begin{equation}
\tan(\Theta^\pm)=\frac{\DSO\pm
2g_\text{orb}\mu_\text{B}B_\parallel}{\DKK}\,.
\label{Theta}
\end{equation}
The corresponding eigenenergies are
\begin{equation}
\boxed{
\begin{split}
&\varepsilon_{1,4}(B_\parallel)=\varepsilon(\Bpar)\pm 
\frac{1}{2}\Delta_\parallel(\Bpar),\\
&\varepsilon_{2,3}(B_\parallel)=\varepsilon(-\Bpar)\pm 
\frac{1}{2}\Delta_\parallel(-\Bpar),
\end{split}}
\label{Eig_en_par}
\end{equation}
where 
$\Delta_\parallel(B_\parallel)=\sqrt{\DKK^2+(\DSO+2g_\text{orb}\mu_\text{B}
B_\parallel)^2}$. Notice the similarity between Eq.~(\ref{Eig_en_par}) and 
Eq.~(\ref{Eig_en_perp}). As in the zero field case, the unitary matrix 
connecting the two bases, Eq.~(\ref{Eig_bas_rel_par}), is block-diagonal in 
spin space. As in the case of the perpendicular magnetic field, the evolution 
in the parallel magnetic field represents rotations by angles $\Theta^\pm/2$ in 
two independent two-dimensional planes, $\Theta^\pm$-planes, involving the 
pairs $(1,4)_P$ and $(2,3)_P$, respectively. The evolution of the energy levels 
is shown in Fig.~\ref{fig:StateEvol}(b). The relevant energy scales 
$\varepsilon(\Bpar), \Delta_\parallel(\Bpar)$ and $\Delta_\parallel(-\Bpar)$ 
are illustrated in Fig.~\ref{fig:elevels}(b). 

As in the case of the perpendicular orientation, one may derive the low and high 
field asymptotics of Eq.~(\ref{Eig_en_par}). Experimentally relevant quantities 
at large magnetic fields, $2g_\text{orb}\mu_\text{B} \Bpar \gg 
\sqrt{\DKK^2+\DSO^2}$, are the energy differences within a Kramers doublet 
$\varepsilon_1(B_\parallel)-\varepsilon_2(B_\parallel)$ and
$\varepsilon_3(B_\parallel)-\varepsilon_4(B_\parallel)$:
\begin{equation}
\begin{split}
&\varepsilon_1(B_\parallel)-\varepsilon_2(B_\parallel)\simeq 
g_\text{s}\mu_\text{B}B_\parallel+\Delta_\text{SO},\\
&\varepsilon_3(B_\parallel)-\varepsilon_4(B_\parallel)\simeq 
-g_\text{s}\mu_\text{B}B_\parallel+\Delta_\text{SO}.
\end{split}
\label{eps1_eps2_high_par}
\end{equation}
Therefore, the relation $[\varepsilon_1(B_\parallel) - 
\varepsilon_2(B_\parallel)] - [\varepsilon_4(B_\parallel) - 
\varepsilon_3(B_\parallel)]\simeq 2\Delta_\text{SO}$ is valid and provides a 
direct way to measure the spin-orbit coupling strength $\Delta_\text{SO}$.
Moreover, the energy difference between chiral pairs is also important. We have
\begin{equation}
\varepsilon_1(\Bpar)-\varepsilon_3(\Bpar)\simeq 
(2g_\text{orb}+g_\text{s})\mu_\text{B}\Bpar,
\end{equation}
\begin{equation}
\varepsilon_2(\Bpar)-\varepsilon_4(\Bpar)\simeq 
(2g_\text{orb}-g_\text{s})\mu_\text{B}\Bpar.
\end{equation}
Thus from the sum $\left[ \varepsilon_1(\Bpar)-\varepsilon_3(\Bpar)\right] 
+\left[ \varepsilon_2(\Bpar)-\varepsilon_4(\Bpar)\right] \simeq 
4g_\text{orb}\mu_\text{B}\Bpar$ the orbital moment can be extracted.

Regarding the low field behavior, we find in leading order in the applied field
\begin{equation}
\begin{split}
&\delta_{\parallel,2}
=\varepsilon_1(B_\parallel)-\varepsilon_2(B_\parallel)\simeq 
g'_+\mu_\text{B}B_\parallel,\\
&\delta_{\parallel,1}
=\varepsilon_3(B_\parallel)-\varepsilon_4(B_\parallel)\simeq 
g'_-\mu_\text{B}B_\parallel,\\
&g'_\pm\equiv\frac{2g_\text{orb}}{\sqrt{1+\bigl(\Delta_\text{KK'}/\Delta_\text{
SO}\bigl)^2}}\pm g_\text{s},
\end{split}
\label{eps1_eps2_low_par}
\end{equation}
which explicitly shows the difference between the effective $g$-factors of the 
two Kramers pairs.

\subsection{Summary}
In Table~\ref{symmTable} we compile the (anti-)commutation relations for the 
different conjugation operations with the different components of 
$\hat{H}_\text{CNT}$, and for $\hat{H}_\text{CNT}$ itself.
\begin{center}
\begin{table}[h]
    \begin{tabular}{| c | c | c | c | c |}
    \hline
    operation & $\hat{H}^{(0)}_\text{CNT}$ & $\hat{H}_\perp$ & 
                           $\hat{H}_\parallel$ & $\hat{H}_\text{CNT}$\\ \hline
    $\hat{\mathcal{T}}$ & commute & anti-comm. & anti-comm. & --- \\ \hline
    $\hat{\mathcal{P}}$ & anti-comm. & anti-comm. & anti-comm. & 
                                                         anti-comm. \\ \hline
    $\hat{\mathcal{C}}$ & anti-comm. & commute & commute & --- \\
    \hline
    \end{tabular}
    \caption{(Anti-)commutation rules of the different combinations of 
operators 
    \{$\hat{\mathcal{T}},\hat{\mathcal{P}},\hat{\mathcal{C}}$\} and 
    \{$\hat{H}^{(0)}_\text{CNT}, \hat{H}_\perp, \hat{H}_\parallel$\} and the 
    resulting (anti-)commutation relations with the total Hamiltonian 
    $\hat{H}_\text{CNT}$ in Eq.~(\ref{Ham_general}).} 
    \label{symmTable}
    \end{table}
\end{center}
The main outcome of our considerations is that both at zero field and for the 
special cases $B_\perp, B_\parallel$  analyzed in this work the spectrum has to 
obey the peculiar conjugation relations 
\begin{equation}
\boxed{
\begin{split}
\hat{\mathcal{T}}-\mathrm{conjugation}\,:\\
\;\varepsilon_{1,4}(-\vec{B}
)=\varepsilon_{2,3}(\vec{B}),
\end{split}
}\label{summary-T}
\end{equation}
\begin{equation}
\boxed{
\begin{split}
\hat{\mathcal{P}}-\mathrm{conjugation}\,:\\
\varepsilon_{1}(\vec{B})-\varepsilon(\vec{B})&=\;\varepsilon(\vec{B}
)-\varepsilon_4(\vec{B})={\Delta(\vec{B})}/{2},\\
\varepsilon_{2}(\vec{B})-\varepsilon(-\vec{B})&=\;\varepsilon(-\vec{B}
)-\varepsilon_3(\vec{B})={\Delta(-\vec{B})}/{2},
\end{split}
}
\label{summary-P}
\end{equation}
where $\Delta_{\perp,\parallel}(\vec{B})$ and $\varepsilon(\vec{B})$ depend 
according to Eqs.~(\ref{Eig_en_perp}) and (\ref{Eig_en_par}) on the modulus and 
direction of the magnetic field. These conjugation relations are dictated by 
the effect of the time-reversal operation as well as by a generalized 
particle-hole operation. All symmetry relations are verified by the 
diagonalization of $\hat{H}_\text{CNT}$. Moreover, it is easy to show using 
Eq.~(\ref{Ham_general}) that Eqs.~(\ref{summary-T}) and (\ref{summary-P}) hold 
true also for an arbitrary direction of the magnetic field $\vec{B}$. The 
conjugation relations impose precise constraints on the form of the Keldysh 
effective action as we shall discuss below. 

\section{Many-particle problem and nonequilibrium field theory}
\label{app:manyparticle}
The experiments presented address to a large extent the effect of the 
electron-electron interaction playing an essential role in the behavior of the 
ultra-clean CNT under investigation. Namely, on top of the nontrivial 
single-particle spectrum controlled by spin-orbit interaction, valley mixing, 
and magnetic field, as discussed in the previous section, the strong electronic 
correlations give rise to a pure many-particle effect, known as the Kondo 
resonance \cite{Hewson_1997}, and lead to a new energy scale $k_\text{B} 
T_\text{K}$ where $T_\text{K}$ is the Kondo temperature. It is this
many-particle state which governs the response of the quantum dot to the 
applied bias voltage $V_\text{sd}$. Since in the experiment this voltage can be
large, the corresponding energy scale $eV_\text{sd}$ can become larger than 
other energy scales and an appropriate nonequilibrium treatment beyond linear 
response is required. Therefore, the experiments challenge the theory which 
must properly take into account the specific single-particle spectrum, in 
particular, its symmetries, electronic correlations and nonequilibrium. Below 
we show the basic concepts of our theory which represents an effective 
field-theoretic approach based on the Keldysh field integral 
\cite{Altland_2010} capable to comprehensively account for different 
single-particle spectra, many-particle interactions and nonequilibrium.

\subsection{Electron-electron interactions in the quantum dot}
Using the states $|i\rangle$, $i=1,2,3,4$, discussed in the previous section
the single-particle Hamiltonian, Eq.~(\ref{Ham_general}), may be written in the
following form
\begin{equation}
\hat{H}_\text{CNT}=\sum_i\varepsilon_i\hat{n}_i,
\label{S_p_H}
\end{equation}
where $\hat{n}_i=d^\dagger_id_i$ and $d^\dagger_i$/$d_i$ are the corresponding
fermionic creation/annihilation operators. Since the electrons in the quantum
dot interact, there is a finite energy cost $U>0$ for two electrons to occupy
the same quantum state. The quantum dot Hamiltonian taking into account the
effect of interactions is
\begin{equation}
\hat{H}_\text{d}=\hat{H}_\text{CNT}+\hat{H}_{ee}.
\label{M_p_H}
\end{equation}
The last term in Eq. (\ref{M_p_H}) describes the electron-electron
interactions in the quantum dot. It has the following form:
\begin{equation}
\hat{H}_{ee}=\frac{U}{2}\sum_{i\neq j}\hat{n}_i\hat{n}_j
\label{ee_Ham}
\end{equation}
and represents one of the key players in the formation of the many-particle
Kondo resonance.

Notice that if $\varepsilon_i=\varepsilon_\text{d}$, $i=1,2,3,4$, the 
Hamiltonian (\ref{M_p_H}) is invariant under $SU(4)$ transformations,
\begin{equation}
\hat{\mathcal{R}}=\exp\bigl(\mathbbmsl{i}\cdot\hat{\mathcal{G}}\bigl),
\label{SU_4}
\end{equation}
where $\hat{\mathcal{G}}$ is an arbitrary Hermitian traceless operator
represented by a four-dimensional matrix in the spin-valley space,
\begin{equation}
\hat{\mathcal{G}}^\dagger=\hat{\mathcal{G}},\quad
\text{Tr}\bigl(\hat{\mathcal{G}}\bigl)=0.
\label{G_op}
\end{equation}
In other words, the electron-electron interaction alone cannot break the
$SU(4)$ symmetry of the system with $\DKK=\DSO=0$.

\subsection{Tunneling between the quantum dot and contacts}
Another essential player responsible for the emergence of the Kondo effect is 
the tunneling coupling between the quantum dot and the conduction electrons in 
the contacts. The electrons in the contacts are assumed to be noninteracting. 
Their Hamiltonian has the form
\begin{equation}
\hat{H}_\text{C}=\sum_{xki}\varepsilon_kc^\dagger_{xki}c_{xki},
\label{C_H}
\end{equation}
where $x$ enumerates the contacts as the left ($x=\text{L}$) and right
($x=\text{R}$) ones, $k$ is the quantum number characterizing the contacts
continuum energy spectrum, $\varepsilon_k$, assumed to be independent of $i$ and
$c^\dagger_{xki}$/$c_{xki}$ are the corresponding creation/annihilation 
operators. The contacts are in equilibrium characterized by chemical potentials
$\mu_\text{L,R}$ such that $\mu_\text{R}-\mu_\text{L}=eV_\text{sd}$.

The electronic exchange between the quantum dot and the contacts is accounted 
for through the tunneling Hamiltonian,
\begin{equation}
\hat{H}_\text{T}=\sum_{xki}(T_{xk}c^\dagger_{xki}d_i+
T^*_{xk}d^\dagger_ic_{xki}),
\label{T_H}
\end{equation}
where $T_{xk}$ are the tunneling matrix elements. Here we assume that the spin 
and orbital degrees of freedom labeled via the index $i$ are conserved during 
the tunneling processes, reflecting the physical situation where the contacts
constitute parts of the same CNT and thus might share the same degrees of 
freedom \cite{Choi2005}. The effect of contacts which may mix orbital quantum 
numbers is thoroughly discussed in Ref.~\onlinecite{Jespersen2011}. There it is 
shown that, despite orbital mixing drives the system from the \SUF\ Kondo fixed 
point to the \SUT\ Kondo fixed point, still \SUF\ Kondo physics governs 
transport for not too large mixing. Tunneling elements preserving orbital 
quantum numbers are considered here for simplicity. However, according to 
Ref.~\onlinecite{Jespersen2011}, also accounting for a small degree of mixing 
should not alter the main conclusion of our work.

If $\varepsilon_i=\varepsilon_\text{d}$, $i=1,2,3,4$, the total Hamiltonian,
\begin{equation}
\hat{H}_\text{tot}\equiv\hat{H}_\text{d}+\hat{H}_\text{C}+\hat{H}_\text{T} 
= \hat{H}_\text{CNT}+\hat{H}_{ee}+\hat{H}_\text{C}+\hat{H}_\text{T},
\label{Ham_tot}
\end{equation}
is still invariant under $SU(4)$ transformations because the tunneling matrix
elements are independent of $i$. In this case the model is usually referred to
as the $SU(4)$ Anderson model \cite{Hewson_1997}. Since in our experiments both
$\DKK$ and $\DSO$ are finite, the $SU(4)$ symmetry is broken.

\subsection{Slave-bosonic transformation}
The Kondo effect studied in our experiments arises in Coulomb valleys with odd
numbers of electrons when the electron-electron interaction $U$ in the quantum
dot significantly exceeds the energy $\Gamma$ (see the definition below)
characterizing the coupling between the quantum dot and contacts. Therefore, to
capture the essence of the Kondo physics it is enough to consider the limit of
strong electron-electron interaction, $U\rightarrow\infty$, when the quantum
dot can accommodate only one electron. In this case the Hamiltonian 
(\ref{M_p_H}) can be diagonalized by means of the so-called slave-bosonic 
transformation \cite{Hewson_1997},
\begin{equation}
d_i=b^\dagger p_i\quad d^\dagger_i=bp^\dagger_i,
\label{S_b}
\end{equation}
where $b^\dagger$/$b$ are bosonic, or slave-bosonic, creation/annihilation 
operators while $p_i^\dagger$/$p_i$ represent new fermionic 
creation/annihilation operators. Physically Eq.~(\ref{S_b}) represents a 
transformation from the electronic states to the states of the quantum dot, 
empty state ($b^\dagger$, $b$) and the state with one electron ($p_i^\dagger$, 
$p_i$). After the diagonalization $\hat{H}_\text{d}$ becomes
\begin{equation}
\hat{H}_\text{d}=\sum_i\varepsilon_ip^\dagger_ip_i.
\label{M_p_H_sb}
\end{equation}

The contacts Hamiltonian is not affected by this transformation while the 
tunneling Hamiltonian becomes
\begin{equation}
\hat{H}_\text{T}=\sum_{xki}\bigl(T_{xk}c^\dagger_{xki}b^\dagger 
p_i+T^*_{xk}bp^\dagger_ic_{xki}\bigl).
\label{T_H_sb}
\end{equation}
As one can see from Eq.~(\ref{T_H_sb}), the slave-bosonic transformation 
simplifying the quantum dot Hamiltonian $\hat{H}_\text{d}$ complicates the 
tunneling Hamiltonian $\hat{H}_\text{T}$ which now, instead of products of two 
second quantized operators, contains products of three second quantized 
operators. The slave-bosonic and new fermionic operators satisfy the constraint
\begin{equation}
b^\dagger b+\sum_i p^\dagger_i p_i=\hat{I},
\label{Constraint}
\end{equation}
which physically reflects the conservation of the total number of the 
slave-bosons and new fermions in the quantum dot which can only have zero or 
one electron.

\subsection{Field integral representation for observables}
The experimental observable of interest is the differential conductance
$G(V_\text{sd})=\partial I/\partial V_\text{sd}$, which can be obtained by 
taking the derivative of the current $I$ through the quantum dot with respect 
to the applied bias voltage $V_\text{sd}$. The current through the quantum dot 
is given by the Meir-Wingreen formula \cite{Wingreen_1994},
\begin{equation}
\begin{split}
I&=\frac{e}{\hbar}\sum_i\int_{-\infty}^\infty 
\!\!\!\!d\varepsilon[n_{\text{R}}(\varepsilon)-n_{\text{L}}(\varepsilon)]\frac{
\Gamma}{4}\frac{W^2}{\varepsilon^2+W^2}\nu_i(\varepsilon),\\
n_\text{L,R}(\varepsilon)&=\frac{1}{\exp[\beta(\varepsilon-\mu_0\pm 
eV_\text{sd}/2)]+1},
\end{split}
\label{MW}
\end{equation}
where $\Gamma\equiv 2\pi\nu_\text{C}|\mathfrak{t}|^2$ ($\nu_\text{C}$ is the 
contacts density of states, $\mathfrak{t}$ is the value of the tunneling matrix 
element $T_{xk}$ assumed to be independent of $x$ and $k$), $W$ is the 
Lorentzian width of the contacts density of states, $\nu_i(\varepsilon)$ is the 
quantum dot tunneling density of states for the state $|i\rangle$, $\beta$ is 
the inverse temperature and $\mu_0$ is the equilibrium chemical potential, 
$\mu_\text{L,R}=\mu_0\mp eV_\text{sd}/2$.

\begin{figure}[t]
\includegraphics[width=\columnwidth]{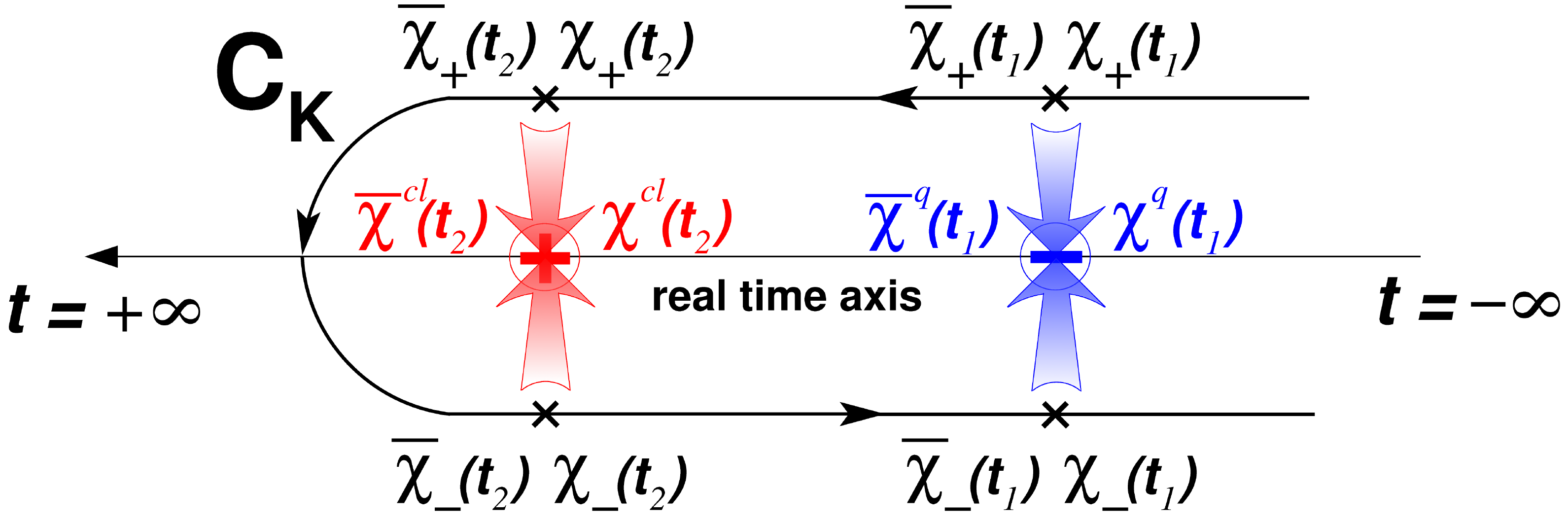}
\caption{\label{Keld_Cont}(Color online) The Keldysh contour $C_\text{K}$, the 
basis of our nonequilibrium theory, is shown. It is used to formulate the real 
time evolution of an interacting nonequilibrium system. Its forward branch goes 
from $t=-\infty$ to $t=+\infty$, its backward branch from $t=+\infty$ to 
$t=-\infty$. The slave-bosonic fields $\bar{\chi},\chi$ living on the forward 
branch are denoted as $\bar{\chi}_+,\chi_+$ while the slave-bosonic fields 
living on the backward branch are denoted as $\bar{\chi}_-,\chi_-$. The forward 
and backward branches are mapped onto the real time axis as sums (red arrows at $t_2$) 
or 
differences (blue arrows at $t_1$) turning, respectively, into the classical or quantum 
slave-bosonic fields via the Keldysh rotation, $\chi^\text{cl,q}(t) \equiv 
(1/\sqrt{2}) [\chi_+(t)\pm\chi_-(t)]$.}
\end{figure}
Therefore, the problem reduces to the calculation of the quantum dot tunneling
density of states,
\begin{equation}
\nu_j(\varepsilon)\equiv-\frac{1}{\pi\hbar}\text{Im}[G^+_{jj}(\varepsilon)],
\label{TDOS_def}
\end{equation}
where $G^+_{jj}(\varepsilon)$ is the quantum dot retarded Green's function for 
the eigenstates $|j\rangle$, $j=1,2,3,4$, of the CNT Hamiltonian. To calculate 
them we develop an effective field theory based on the Keldysh field integral
\cite{Altland_2010} where one replaces all the second quantized operators by the
corresponding fields. The basic idea of this theory is to use the advantage of 
the physical clarity provided by the slave-bosonic transformation, introduced 
in the previous subsection, which allows one to deal directly with the states 
of the quantum dot. In particular, in the Kondo regime the probability of the 
empty state of the quantum dot is small and, thus, large fluctuations of the 
slave-bosonic fields, describing the empty state, are not relevant for the 
Kondo physics. One can then use a low order expansion in these fields around 
proper field configurations in order to calculate the quantum dot observables.

The practical implementation of this idea involves a functional integration over
all the fermionic, or Grassmann, fields, describing the electrons in contacts 
and the states of the quantum dot with one electron. After this step one is 
left with an effective field theory describing the dynamics of the 
slave-bosonic field. At this stage any quantum dot observable,
$\hat{O}=\mathcal{F}(d_\sigma^\dagger$, $d_\sigma)$, originally expressed in 
terms of the second quantized operators, admits a field-theoretic 
representation based on the Keldysh effective action \cite{Smirnov_2011, 
Smirnov_2011a, Smirnov_2013}:
\begin{equation}
\begin{split}
\langle\hat{O}\rangle(t)= \frac{1}{\mathcal{N}_0}\underset{\mu\rightarrow\infty}
{\text{lim}}e^{\beta\mu} &
\int\mathcal{D}[\bar{\chi},\chi]e^{\frac{\mathbbmsl{i}}{\hbar}S_\text{eff}[\bar{
\chi}^\text{cl,q}(\tilde{t});\chi^\text{cl,q}(\tilde{t})]} \\
&\times\mathcal{F}[\bar{\chi}^\text{cl,q}(t);\chi^\text{cl,q}(t)],\\[2mm]
S_\text{eff}[\bar{\chi}^\text{cl,q}(t);\chi^\text{cl,q}(t)]= & S_0[\bar{\chi}
^\text{cl,q}(t);\chi^\text{cl,q}(t)] \\
&+ S_\text{T}[\bar{\chi}^\text{cl,q}(t);\chi^\text{cl,q}(t)],
\end{split}
\label{KFI_observ}
\end{equation}
where $\chi^\text{cl,q}$ are the classical and quantum \cite{Altland_2010}
eigenstates of the bosonic annihilation operator $b$, $\mathcal{N}_0$ is a
normalization constant \cite{Smirnov_2011} and the limit $\mu\rightarrow\infty$ 
in Eq.~(\ref{KFI_observ}) takes into account the constraint in 
Eq.~(\ref{Constraint}). The Keldysh effective action, $S_\text{eff} 
[\bar{\chi}^\text{cl,q}(t); \chi^\text{cl,q}(t)]$, in Eq.~(\ref{KFI_observ}) is 
the sum of $S_0[\bar{\chi}^\text{cl,q}(t);\chi^\text{cl,q}(t)]$, being the 
standard free bosonic action \cite{Altland_2010} on the Keldysh contour, Fig. 
\ref{Keld_Cont}, and $S_\text{T}[\bar{\chi}^\text{cl,q}(t); 
\chi^\text{cl,q}(t)]$, being the tunneling action of the problem.

\subsection{Keldysh effective action and tunneling density of states}
The tunneling term of the Keldysh effective action,
\begin{equation}
S_{\rm T}[\bar{\chi}^{\rm cl,q}(t);\chi^{\rm cl,q}(t)]=-\mathbbmsl{i}\hbar\,{\rm 
tr}\ln\bigl[-\mathbbmsl{i}G^{(0)-1}-\mathbbmsl{i}\mathcal{T}\bigl],
\label{tun_act}
\end{equation}
is a highly nonlinear functional of the slave-bosonic fields. Here the matrix
$\mathcal{T}$ is off-diagonal in the quantum dot-contacts space,
\begin{equation}
\mathcal{T}=
\begin{pmatrix}
0 & M_{\rm T}^\dagger(i t|k'i't')\\
M_{\rm T}(ki t|i't') & 0
\end{pmatrix},
\label{T_matr}
\end{equation}
\begin{equation}
\begin{split}
M_{\rm T}(ki t|i' t')=&\frac{\delta(t-t')\delta_{ii'}\mathfrak{t}}{\sqrt{2}\hbar} \\
&\times
\begin{pmatrix}
\bar{\chi}^{\rm cl}(t)-\gamma_i\sqrt{2} & \bar{\chi}^{\rm q}(t)\\
\bar{\chi}^{\rm q}(t) & \bar{\chi}^{\rm cl}(t)-\gamma_i\sqrt{2}
\end{pmatrix},
\label{tun_matr}
\end{split}
\end{equation}
\begin{equation}
\begin{split}
M^\dagger_{\rm T}(i t|k'i' 
t')=&\frac{\delta(t-t')\delta_{ii'}\mathfrak{t}^*}{\sqrt{2}\hbar}\\
&\times
\begin{pmatrix}
\chi^{\rm cl}(t)-\delta_i\sqrt{2} & \chi^{\rm q}(t)\\
\chi^{\rm q}(t) & \chi^{\rm cl}(t)-\delta_i\sqrt{2}
\end{pmatrix}.
\end{split}
\label{tun_matr_conj}
\end{equation}
In Eqs.~(\ref{tun_matr}) and (\ref{tun_matr_conj}) $\gamma_i$ and $\delta_i$ 
($i=1,2,3,4$) represent the initially arbitrary expansion points or shifts of 
the classical slave-bosonic fields, $\bar{\chi}^\text{cl}$ and 
$\chi^\text{cl}$, respectively, in the slave-bosonic space. As shown below, 
$\gamma_i$ and $\delta_i$ can be determined from the symmetries of the 
Hamiltonian and from the Fermi-liquid behavior at zero temperature and zero 
bias. Notice, that since the integration variables $\bar{\chi}^\text{cl,q}$ and 
$\chi^\text{cl,q}$ are independent of each other, the $\gamma_i$ and $\delta_i$ 
are not complex conjugates.

The Green's function matrix $G^{(0)}$ is block-diagonal in the quantum 
dot-contacts space. Its quantum dot block $G^{(0)}_{\rm d}$ has the standard 
$2\times 2$ fermionic Keldysh structure:
\begin{equation}
G^{(0)}_{\rm d}(i t|i' t')=\delta_{ii'}
\begin{pmatrix}
G^+_i(t-t') & G^{\rm K}_i(t-t')\\
0 & G^-_i(t-t')
\end{pmatrix}.
\label{G_0_matr}
\end{equation}
In the frequency domain the components of the above matrix are
\pagebreak
\begin{equation}
\begin{split}
&G^+_i(\omega)=\frac{\hbar}{\hbar\omega-(\varepsilon_i+\mu)+\mathbbmsl{i}E_{i}},
\quad
G^-_i(\omega)=[G^+_i(\omega)]^*,\\
&G^{\rm 
K}_i(\omega)=\frac{1}{2}[G^+_i(\omega)-G^-_i(\omega)]\sum_x\tanh\biggl[\frac{
\hbar\omega-\mu_x}{2k_\text{B}T}\biggl].
\end{split}
\label{G_0_matr_como_fr}
\end{equation}
Here $E_{i}\equiv\gamma_{i}\delta_{i}\Gamma/2$.

In the case of the quantum dot tunneling density of states, $\nu_j$, for the 
state $|j\rangle$ the expression for the integrand in Eq.~(\ref{KFI_observ}) is
\begin{widetext}
\begin{equation}
\begin{split}
\mathcal{F}[\bar{\chi}^{\rm cl,q}(t);\chi^{\rm 
cl,q}(t)]=[\bar{\chi}_-(t)\chi_+(0)-\bar{\chi}_+(t)\chi_-(0)]
&\times[G^{(0)-1}+\mathcal{T}]^{-1}(j t|j 0),
\end{split}
\label{F_TDOS}
\end{equation}
\end{widetext}
where $\bar{\chi}_{\pm}$, $\chi_{\pm}$ are the slave-bosonic fields on the 
forward and backward branches \cite{Altland_2010} of the Keldysh contour,
Fig.~\ref{Keld_Cont}, and the expansion points, introduced in 
Eqs.~(\ref{tun_matr}) and (\ref{tun_matr_conj}), are labeled by an additional 
upper index, $\gamma_i,\delta_i \rightarrow \gamma^j_i,\delta^j_i$ and, as a 
consequence $E_i\rightarrow E^j_i$.

As mentioned in the previous subsection, to solve this highly nonlinear problem
one has to perform an expansion of the tunneling action in powers of the
slave-bosonic fields. To get the relevant physics already in the lowest,
{\it i.e.}, in the second order expansion, one must carefully specify the field
configurations around which this expansion has to be performed. This can be 
done with a suitable choice of the expansion points $\gamma^j_i$ and 
$\delta^j_i$. Since the linear terms in the Keldysh effective action do not 
generate (see Ref. \onlinecite{Smirnov_2013}) any finite contribution to 
$\nu_j(\varepsilon)$, in the second order expansion the expansion points 
$\gamma^j_i$ and $\delta^j_i$ appear only through $E^j_{i}$ which determine the 
form of the propagators in Eq.~(\ref{G_0_matr_como_fr}). Therefore, 
$\gamma^j_i$ and $\delta^j_i$ just renormalize the kernel of the quadratic 
Keldysh effective action.

To properly account for the Kondo correlations specific to CNT quantum dots this
renormalization should respect the specific symmetry properties of the energy
spectrum. First, since TR symmetry is broken at finite magnetic fields, the 
Keldysh effective action should also reflect this symmetry breaking together 
with the $\hat{\mathcal{T}}$-conjugation, Eq.~(\ref{summary-T}). Second, the 
$\hat{\mathcal{P}}$-conjugation relations, Eq.~(\ref{summary-P}), are valid 
at any magnetic field and it is natural to construct a Keldysh effective action 
having these properties as well. It is easy to see that these two requirements 
lead to a structure of the Keldysh effective action where only two states from 
different Kramers pairs and from the same particle-hole pair are present in the 
action.

Finally, to identify which particle-hole pair should be retained in the Keldysh
effective action, we recall that the Kondo effect in quantum dots, whose states
are characterized by a discrete quantum number $i$, arises from virtual
transitions between the electronic states $|i\rangle$ and $|i'\rangle$. Even 
though both flip $i\neq i'$ and non flip terms $i=i'$ are important for the 
Kondo resonance, non flip processes alone cannot give rise to the Kondo effect, 
as seen for example from the analysis of the renormalization group flow 
equations for the \SUF\ Kondo effect in carbon nanotubes \cite{Choi2005}. This 
is similar to the $SU(2)$ Anderson model, where the spin-$1/2$ Kondo behavior 
arises from the virtual transitions between the electronic states 
$|\uparrow\rangle$ and $|\downarrow\rangle$ in the quantum dot, which are, in 
this case, spin flips. Therefore, to properly capture the Kondo behavior 
already in the lowest order expansion it is natural, in view of the poor man's 
scaling where the renormalization flow is governed by flip processes, in the 
calculation of $\nu_j(\varepsilon)$ to effectively eliminate from the kernel 
of the Keldysh effective action the propagator for the same state $|j\rangle$ 
(see Eq. (\ref{G_0_matr_como_fr})).

The above considerations {\it uniquely} specify the structure of the Keldysh
effective action. To obtain this structure one has to choose the expansion 
points $\gamma_i^j$ and $\delta_i^j$ appropriately. Let us for example discuss 
how we choose $E^1_{i}=\gamma_i^1\delta_i^1\Gamma/2$ in the Keldysh effective 
action for $\nu_1(\varepsilon)$. Since the Keldysh effective action for 
$\nu_1(\varepsilon)$ should effectively contain the pair of propagators for the 
states $|2\rangle$ and $|3\rangle$ only, the expansion points are chosen such 
that the real and imaginary parts of $E^1_{i}$ are
\begin{equation}
\begin{split}
E^{1\text{R}}_{1}&=E^{1\text{R}}_{2}=E^{1\text{R}}_{3}=E^{1\text{R}}_{4}=E^\text
{
R},\\
E^{1\text{I}}_{1}&=E^\text{I}-(\varepsilon_1-\varepsilon_2),
\\
E^{1\text{I}}_{2}&=E^\text{I},
\\
E^{1\text{I}}_{3}&=E^\text{I},
\\
E^{1\text{I}}_{4}&=E^\text{I}-(\varepsilon_4-\varepsilon_3), 
\end{split}
\label{expansionpoints}
\end{equation}
where $E^\text{R,I}$ are currently arbitrary but their values may be found from 
the Fermi-liquid behavior (see below). In a similar way the parameters 
$E^2_{i}, E^3_{i}$, and $E^4_{i}$ entering the calculation of 
$\nu_2(\varepsilon), \nu_3(\varepsilon)$ and $\nu_4(\varepsilon)$, respectively,
can be chosen.

Since the Keldysh effective action is quadratic, the final expressions for
the tunneling densities of states are obtained by calculating the corresponding 
Gaussian field integrals. We find
\begin{widetext}
\begin{equation}
\nu_j(\varepsilon,\vec{B})=\frac{1}{2\pi}\frac{W^2}{\varepsilon^2+W^2}
\frac{\Gamma}{[\varepsilon_j(\vec{B}) - \varepsilon + 
\Gamma\Sigma^\text{R}_j(\varepsilon,\vec{B})]^2 + 
[\Gamma\Sigma^\text{I}_j(\varepsilon,\vec{B})]^2}.
\label{TDOS}
\end{equation}
In Eq.~(\ref{TDOS}) $\Sigma^\text{R}_j(\varepsilon,\vec{B})$ and 
$\Sigma^\text{I}_j(\varepsilon,\vec{B})$ are, respectively, the real and
imaginary parts of the self-energies,
\begin{equation}
\Sigma_j(\varepsilon,\vec{B})=-\sum_{i=1}^4\frac{1}{2}\int_{-\infty}^\infty\frac
{d\varepsilon'}{2\pi}\frac{W^2}{\varepsilon'^2+W^2}
\frac{n_\text{L}(\varepsilon')+n_\text{R}(\varepsilon')}{
\varepsilon'-\varepsilon-(\varepsilon_i(\vec{B})-\varepsilon_j(\vec{B}
))+\mathbbmsl{i}E^j_i}.
\label{S_e1}
\end{equation}
With the $E^j_i$ chosen as discussed in Eq.~(\ref{expansionpoints}) this turns 
into
\begin{equation}
\begin{split}
&\Sigma_{1,4}(\varepsilon,\vec{B})=-\sum_{i=2,3}\int_{-\infty}^\infty\frac{
d\varepsilon'}{2\pi}\frac{W^2}{\varepsilon'^2+W^2}
\frac{n_\text{L}(\varepsilon')+n_\text{R}(\varepsilon')}{
\varepsilon'-\varepsilon-(\varepsilon_i(\vec{B})-\varepsilon_{1,4}(\vec{B}
))+\mathbbmsl{i}E},\\
&\Sigma_{2,3}(\varepsilon,\vec{B})=-\sum_{i=1,4}\int_{-\infty}^\infty\frac{
d\varepsilon'}{2\pi}\frac{W^2}{\varepsilon'^2+W^2}
\frac{n_\text{L}(\varepsilon')+n_\text{R}(\varepsilon')}{
\varepsilon'-\varepsilon-(\varepsilon_i(\vec{B})-\varepsilon_{2,3}(\vec{B}
))+\mathbbmsl{i}E},
\end{split}
\label{S_e2}
\end{equation}
\end{widetext}
where virtual non flip processes and $\hat{\mathcal{P}}$-processes no longer 
explicitly appear. It is easy to see from Eq.~(\ref{S_e2}) that, as a result of 
the time-reversal and particle-hole conjugation relations of the single 
particle energies $\varepsilon_j(\vec{B})$, also the so obtained observables 
$\nu_j(\varepsilon)$ (here we explicitly indicate that $\nu_j$ depend on the 
magnetic field $\vec{B}$ or on the splittings $\Delta(\pm\vec{B})$) fulfill the 
conjugation relations
\begin{equation}
\nu_1(\varepsilon,\vec{B})=\nu_2(\varepsilon,-\vec{B}),\quad\nu_3(\varepsilon,
\vec{B})=\nu_4(\varepsilon,-\vec{B}),
\end{equation}
\begin{equation}
\begin{split}
\nu_{1}(\varepsilon,\Delta(\vec{B}))=\nu_{4}(\varepsilon,-\Delta(\vec{B})),\\
\nu_{2}(\varepsilon,\Delta(-\vec{B}))=\nu_{3}(\varepsilon,-\Delta(-\vec{B})),
\end{split}
\end{equation}
where $\Delta(\vec{B})\equiv\varepsilon_1(\vec{B})-\varepsilon_4(\vec{B})$
and $\Delta(-\vec{B})\equiv\varepsilon_2(\vec{B})-\varepsilon_3(\vec{B})$ 
according to Eq.~(\ref{summary-P}).

At zero temperature, zero bias and zero magnetic field in Coulomb valleys with 
odd numbers of electrons the differential conductance reaches its unitary limit 
value $2e^2/h$, as predicted by the Fermi-liquid theory \cite{Hewson_1997}. The 
tunneling density of states has a narrow maximum located close to the 
equilibrium chemical potential $\mu_0$. Since in our experiments $\Delta \equiv 
\sqrt{\Delta^2_\text{KK'}+\Delta^2_\text{SO}}\simeq 7k_\text{B}T_\text{K}$, one 
can neglect a small deviation from $\mu_0$ in the location of this maximum. 
Using the two conditions, the unitary limit value and location of the maximum, 
one can obtain the real and imaginary parts $E^\text{R}$ and $E^\text{I}$. 
After that the expressions for the tunneling densities of states, 
Eq.~(\ref{TDOS}), can be used to calculate the differential conductance 
$G(V_\text{sd},T,B)$ as a function of the bias voltage, temperature, and 
magnetic field.

Note that in principle one might use knowledge of the unitary limit for another 
observable like the magnetic susceptibility (see 
Ref.~\onlinecite{Hewson_1997}), which would not change $E^\text{R,I}$ since the 
unitary limits for different observables are related.

\subsection{The Kondo temperature and universality}
\begin{figure}
\includegraphics[width=7cm]{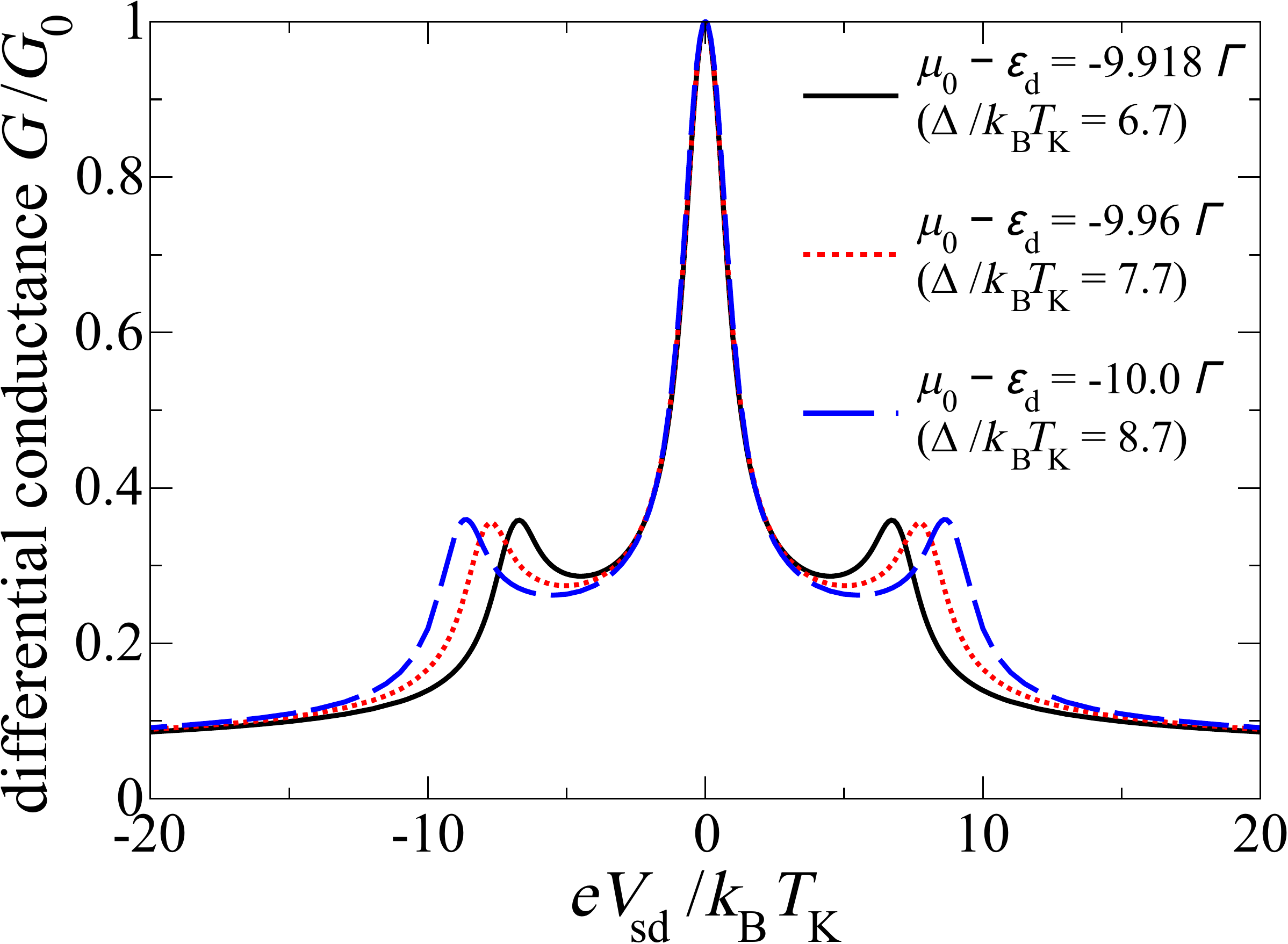}
\caption{\label{figure_3_supp}(Color online) The differential conductance as a 
function of the bias voltage at zero temperature and zero magnetic field for 
different values of $\mu_0-\varepsilon_d$ (equivalently for different gate 
voltages) and for a fixed value of $\Delta$. The Kondo temperature 
$T_\text{K}^{(0)}$ is here rescaled to $T_\text{K}$ which is the temperature at 
which the linear conductance as a function of the temperature is equal to 
$G_0/2$. In our model $G_0=2e^2/h$, since the couplings to the left and right 
contacts are assumed to be equal. As one can see, the zero bias peak, or the 
Kondo peak, is universal while the side peaks are not. The full universality, 
{\it i.e.}, the universality at any bias voltage is recovered only if $\Delta / 
k_\text{B} T_\text{K} = \text{const}$ which is not the case in our experiments 
because the gate voltage is altered for a fixed CNT quantum dot, that is for a 
fixed value of $\Delta$.}
\end{figure}
The theory presented above predicts the universal behavior of the differential 
conductance with the scaling given by the Kondo temperature
\begin{equation}
k_\text{B}T_\text{K}^{(0)}=f\biggl[\frac{\Delta}{k_\text{B}T_\text{K}^{SU(4)}}
\biggl]2W\exp\biggl[-\frac{\mu_0-\varepsilon_d}{2\Gamma}\biggl]
\label{Kondo_temp}
\end{equation}
(differing from $k_\text{B}T_\text{K}$, used in the main text, by a constant
prefactor, chosen such that the linear conductance at $T_\text{K}$ equals
half of its unitary value). In Eq.~(\ref{Kondo_temp})
\begin{equation}
k_\text{B}T_\text{K}^{SU(4)}=2W\exp\biggl[-\frac{\mu_0-\varepsilon_d}{2\Gamma}
\biggl]
\label{Kondo_temp_SU4}
\end{equation}
is the Kondo temperature for the $SU(4)$ Anderson model and $f(x)$ is a 
function of the ratio $x=\Delta/k_\text{B}T_\text{K}^{SU(4)}$. Our theory
correctly predicts the limiting behavior of this function. When
$\Delta\rightarrow\infty$ ($x\rightarrow\infty$) we find that
\begin{equation}
f(x)\rightarrow \exp\biggl[-\frac{\mu_0-\varepsilon_d}{2\Gamma}\biggl].
\label{Limit_f_Delta_inf}
\end{equation}
This leads to
\begin{equation}
T_\text{K}^{(0)}\rightarrow 
T^{SU(2)}_\text{K}=2W\exp\biggl[-\frac{\mu_0-\varepsilon_d}{\Gamma}\biggl].
\label{Limit_TK_Delta_inf}
\end{equation}
On the other side, when $\Delta\rightarrow 0$ ($x\rightarrow 0$), we find that
\begin{equation}
f(x)\rightarrow 1
\label{Limit_f_Delta_0}
\end{equation}
leading to
\begin{equation}
T_\text{K}^{(0)}\rightarrow T^{SU(4)}_\text{K}.
\label{Limit_TK_Delta_0}
\end{equation}
The correct scaling behavior for any $\Delta$ is one of the advantages of our 
theory over other theories, {\it e.g.}, over the equations of motion technique 
which is a popular tool to investigate the Kondo effect in various setups, in 
particular, in CNT quantum dots \cite{Fang_2008}.

It is essential that the Kondo correlations significantly weaken the loss of 
universality at energies smaller than $\Delta$. As a result, the central peak 
remains almost universal: the variation in its width is much smaller than the 
relative variation (about 30\% in Fig.~\ref{figure_3_supp}) of the Kondo energy 
scale \kTK. This behavior is specific to the \SUF\ broken Kondo effect and
represents its fingerprint when observed in experiments.

\begin{figure*}[t]
\begin{center}
\includegraphics[width=12.5cm]{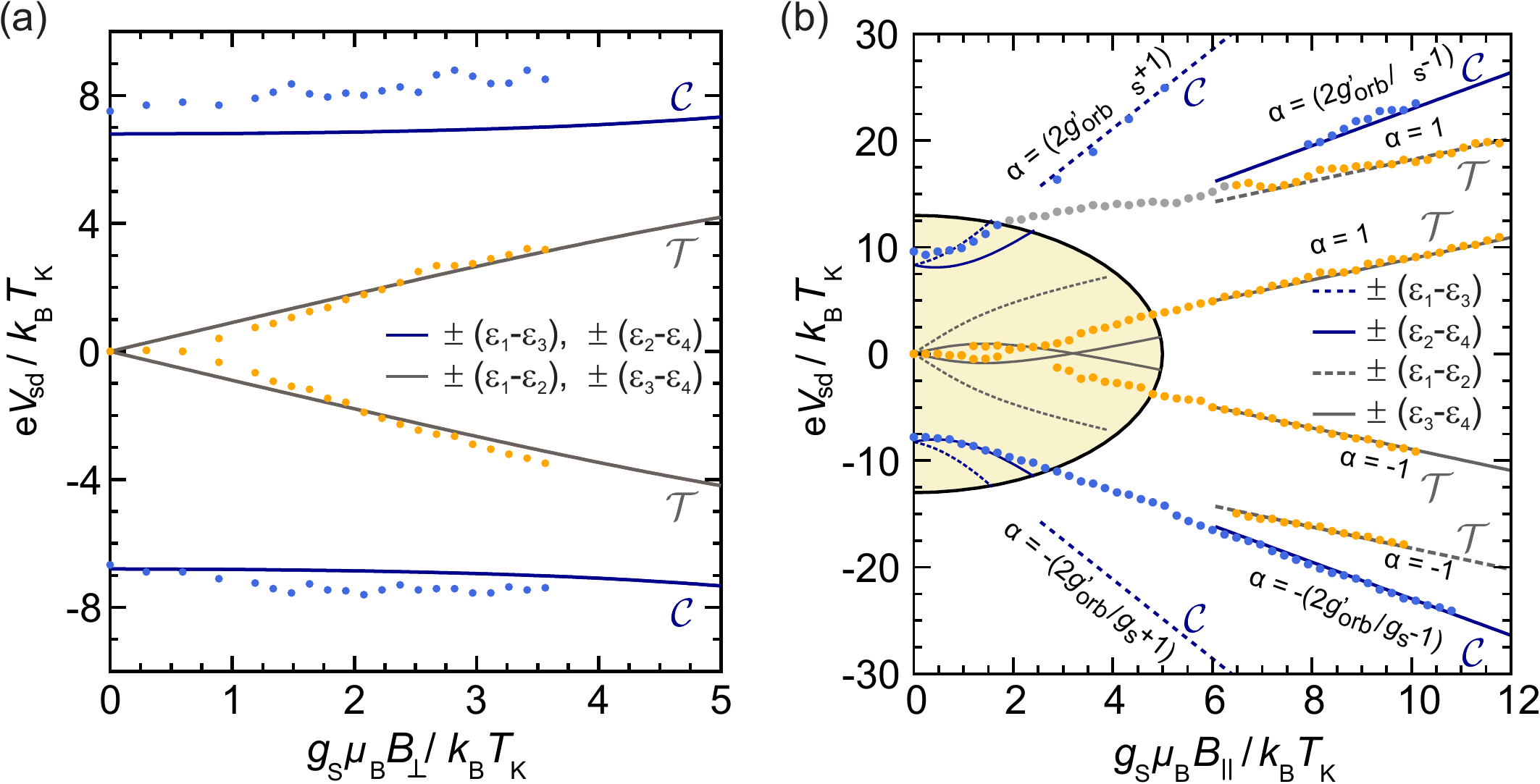}
\end{center}
\caption{
(Color) Calculated single particle energy differences between CNT quantum 
levels (lines) and corresponding measured positions of conductance peaks (dots) 
for a wide range of bias voltages and magnetic fields perpendicular (a) and 
parallel (b) to the CNT axis. In each case, gray lines correspond to 
intra-Kramers and blue lines to chiral inter-Kramers transitions. For the case 
of parallel magnetic field (b), a discussion of low field (yellow area) versus 
high field behaviour can be found in the text. In the high field limit, the 
$\alpha$-values describe the slopes of the linear magnetic field dependence.
}
\label{fig:suppcomparison}
\end{figure*}
\section{Inelastic conductance peaks at high magnetic fields}
\label{app:inelastic}
In this section we discuss in more detail the evolution of the measured 
conductance peaks in a perpendicular and parallel magnetic field, as shown 
in Fig.~\ref{fig:suppcomparison} (see also Fig.~\ref{fig:comparison}).
The evolution of the satellite peak positions with magnetic field is governed 
by the energy differences $\varepsilon_1-\varepsilon_3$ and $\varepsilon_2 - 
\varepsilon_4$ (blue dashed and solid lines). In perpendicular field these 
differences weakly depend on the magnitude of the magnetic field while they 
strongly vary in parallel field. Hence, we expect that the location and height 
of the satellite peaks will not be affected by the perpendicular field \Bperp\ 
but must strongly depend on \Bpar. Indeed, this is what we observe in
Fig.~\ref{fig:suppcomparison}(a) and Fig.~\ref{fig:suppcomparison}(b), 
respectively. According to our theory, increasing the magnitude of \Bpar\ 
shifts the satellite peaks to higher voltages. At $\Bpar\gtrsim 
1.1\,\kTK/g_\text{s}\muB$, the satellite peaks split into two peaks (see e.g. 
the red curve in Fig.~\ref{fig:magnetic}(f)). By further increasing \Bpar, the 
peak splitting grows while both peaks continue to move to higher voltages. This 
behavior is again in a good qualitative agreement with our experiments, albeit 
the second peak splitting occurs at unexpectedly low magnetic field.

In the case of the perpendicular field, Fig.~\ref{fig:suppcomparison}(a), the 
peak positions related to intra-Kramers and chiral inter-Kramers transitions 
match the expectations from the theory in the whole range of the investigated 
magnetic field values. On the other hand, in a parallel field one can
clearly distinguish between a low field region [contained in the yellow area in 
Fig.~\ref{fig:suppcomparison}(b)] and a high field region. The latter shows a 
linear dependence both of the peak positions and of the energy differences on 
the magnetic field. The low field region is defined by the condition $\Bpar \ll 
\Delta/2g_\text{orb}\mu_\text{B}$, where $g_\text{orb}\mu_\text{B}$ is the 
orbital magnetic moment (see Eq.~(\ref{Ham_general})). With the values of 
\begin{table}[b]
\begin{tabular}{ |c | c | c || c | c | c |}
\hline
\textbf{perpend.} & theory & experiment & \textbf{parallel} & theory & 
experiment \\ \hline \hline
\TK\,(K)     &    & 0.86 & \TK\,(K)      &    & 1.12    \\ \hline
$\Delta$     & 6.9  & 6.8, 7.5 & $\Delta$      & 8.3   & 7.5, 10 \\ 
\hline
$g_\text{s}$ &  2 &      & $g_\text{s}$  &  2 &         \\ \hline
$g'_\text{s}$& 1.80   &$1.85\pm 0.05$ & $g_\text{orb}/g_\text{s}$ &  
2.4 &  \\ \hline
&    &      & $g'_\text{orb}/g_\text{s}$       & 1.37 & 
1.37 \\ \hline
\DSO     & 3.0 &  & \DSO   & 3.6  & 4.7  \\ \hline
\DKK    & 6.2   &  & \DKK   & 7.5 &   \\ \hline
\end{tabular}
\caption{
Values of the nanotube parameters as used in the theoretical calculations 
and extracted from the differential conductance measurements in magnetic fields 
perpendicular (valley with $\nel=21$) and parallel (valley with $\nel=17$) to 
the nanotube axis. Energies are given in units of \kTK. The two values of 
$\Delta$ given in the experimental columns correspond to the values extracted 
from the satellite peaks in the differential conductance at negative and 
positive bias voltages, respectively.}
\label{values}
\end{table}
$\Delta/\kTK$ and $g_\text{orb}$ (see Table~\ref{values}), used in the 
numerical calculations, this corresponds to $\Bpar \ll 1.5\,\kTK / g_\text{s} 
\muB$, where $g_\text{s}$ is the spin $g$-factor. In the high field region, 
$\Bpar \gg \Delta/2g_\text{orb}\mu_\text{B}$, the valley mixing effects become 
negligible and the energy eigenstates are products of pure spin and valley 
states. In this case the data can be fitted to the linearized dispersion 
relation $\varepsilon_{1,4}(\Bpar) \propto \muB \Bpar (g_\text{s}/2 \pm 
g_\text{orb}')$ and $\varepsilon_{2,3}(\Bpar) \propto \muB \Bpar (-g_\text{s}/2 
\pm g_\text{orb}')$, with $g_\text{orb}'/g_\text{s}=1.37$. 
\begin{figure*}[t]
\includegraphics[width=\textwidth]{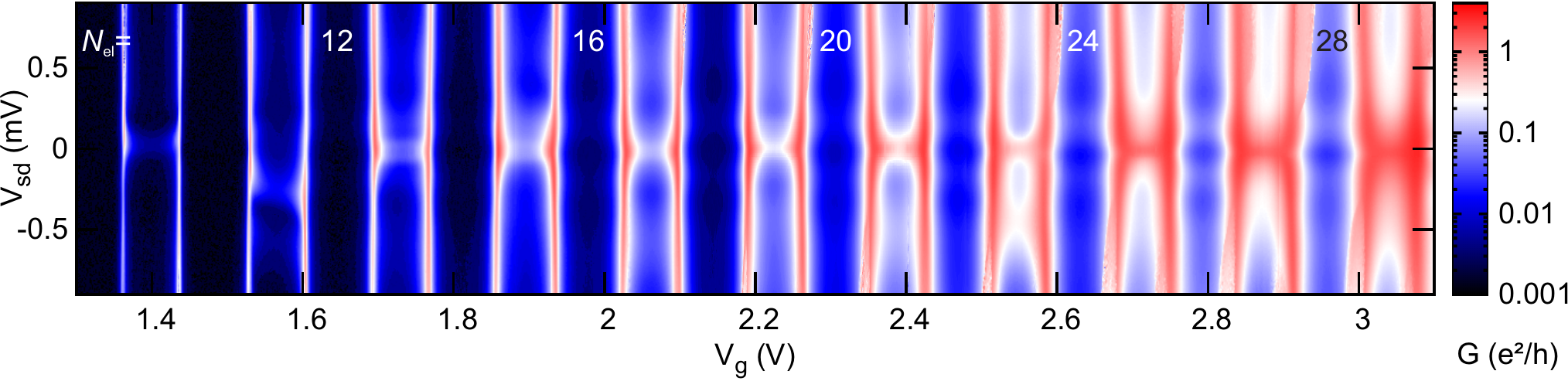}
\caption{(Color) Differential conductance $G$ as function of bias voltage \vsd\ 
and gate voltage \vg, for the gate voltage region corresponding to $9\le \nel 
\le 29$.}
\label{diamonds}
\end{figure*}
\begin{figure*}[t]
\includegraphics[width=\textwidth]{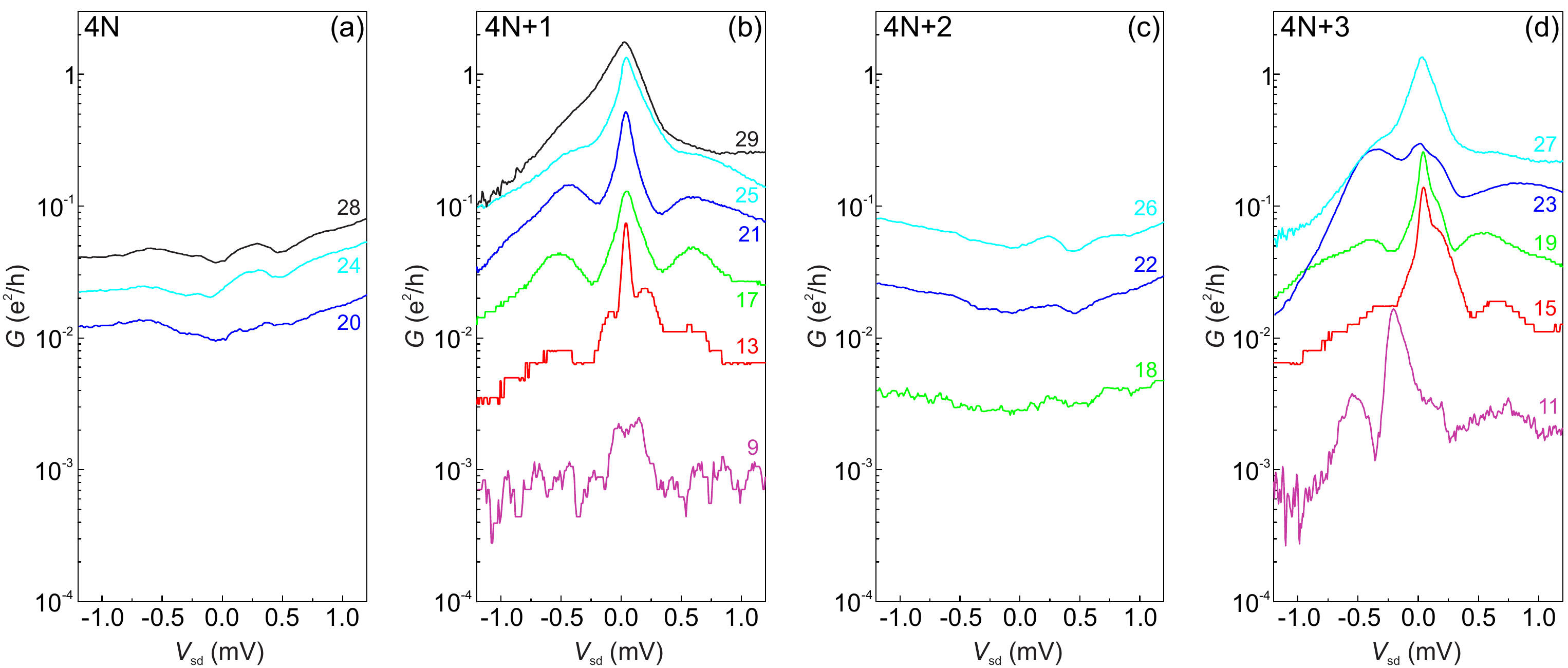}
\caption{(Color online) Differential conductance $G$ as function of bias 
voltage \vsd, at gate voltages \vg\ in the center of the Coulomb blockade 
valleys for differing electron numbers. The four panels each display traces 
corresponding to a particular residue modulo 4 of the electron number \nel.}
\label{shells}
\end{figure*}
\begin{figure}[t]
\includegraphics[width=0.25\textwidth]{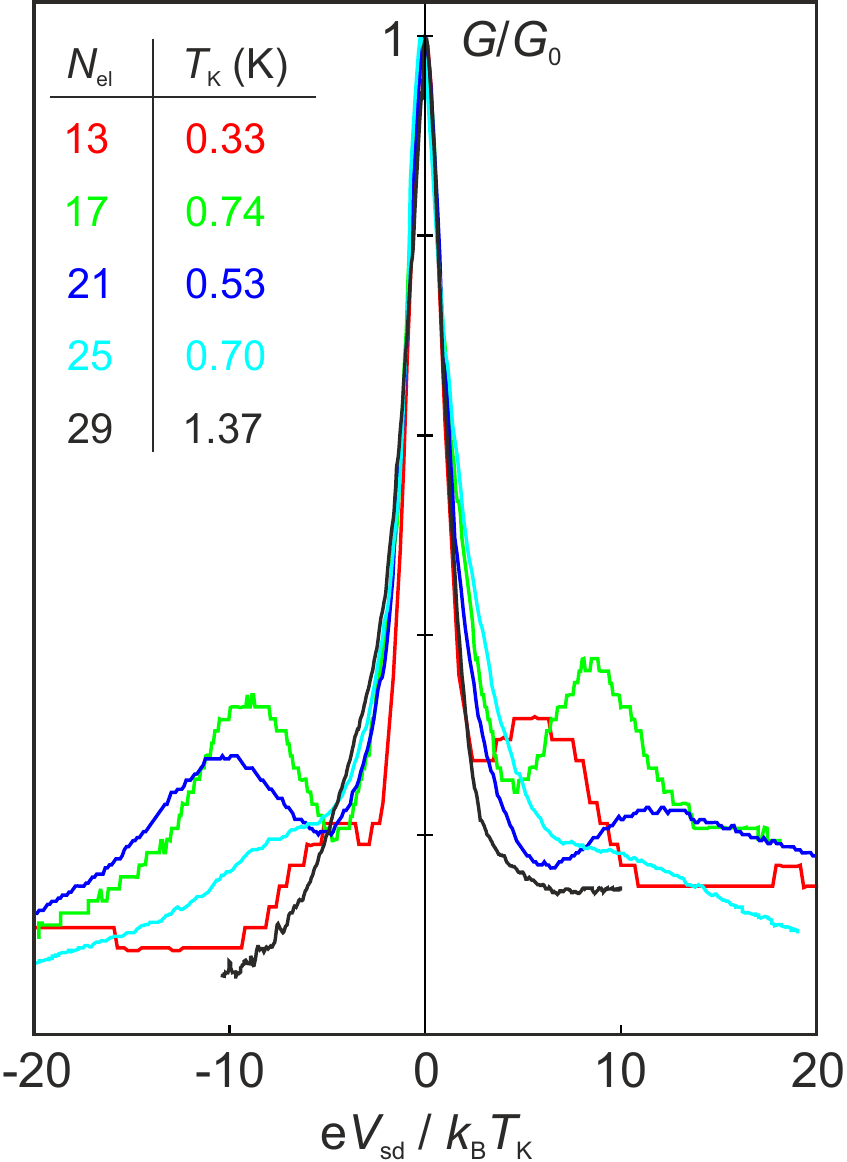}
\caption{(Color online) Line traces $G(\vsd)$ from Fig.~\ref{shells}(b), i.e., 
at the center of Coulomb blockade valleys with occupation $4N+1$, with \vsd\ 
rescaled by the corresponding Kondo temperature $\TK$ and the conductance 
rescaled by its maximum value $G_0=G(\vsd=0)$.}
\label{shellsrescaled}
\end{figure}
This amounts to a 
ratio between high field and low field orbital g-factors of $g_\text{orb}' / 
g_\text{orb} = 0.57$. Notice that such asymptotic behavior is also predicted by 
expanding the spectrum of the effective Hamiltonian $\hat{H}_\text{CNT}$ at 
large fields, see Eqs.~(\ref{eps1_eps2_high_par}), (\ref{eps1_eps2_low_par}), 
but there $g_\text{orb}'/g_\text{orb}=1$. This suggests the presence of a 
mechanism which reduces the ratio of the orbital $g$-factors that is not 
captured by the Hamiltonian $\hat{H}_\text{CNT}$ and goes beyond the scope of 
this work. This is the reason why only a qualitative but not quantitative 
agreement between theory and experiment in the investigated range of applied 
parallel fields is found. Accounting for this discrepancy, it is possible to 
identify all Kondo transitions observed in the experiment as either 
intra-Kramers or chiral inter-Kramers transitions. In particular 
intra-Kramers transitions involving the excited Kramers doublet $(1,2)_T$ 
become visible at larger fields (see dashed gray line 
in Fig.~\ref{fig:suppcomparison}(b)), while they are not discernible at small 
magnitudes of the parallel magnetic field.

Finally, in the linear regime $\Bperp\gtrsim 5.3\,B_c$ it is possible to 
experimentally extract the effective $g$-factor $g'_\text{s} = g_\text{s} / 
\sqrt{1+(\DSO/\DKK)^2}$ (see Eq.~(\ref{eps1_eps_2_perp_low})), and hence 
to directly access the ratio $\DSO/\DKK$ for the valley $\nel=21$. Similarly, 
from the linear regime $\Bpar\gtrsim 1.75\,B_\text{c3}$ we can directly extract 
\DSO\ (see Eq.~(\ref{eps1_eps2_high_par})) for the valley $\nel=17$. These 
values together with the parameters used in the numerical calculations are 
shown in Table~\ref{values}.

\section{Systematic evolution of the transport spectrum with the electron 
number}
To demonstrate the generic character of the observed fine structure of the 
Kondo resonance in charging states with odd electron number, we present a more 
comprehensive overview over the different Coulomb valleys for electron numbers 
between 9 and 29. 

Fig.~\ref{diamonds} displays the stability diagram measurement, i.e. the 
differential conductance $G$ as function of bias voltage \vsd\ and gate voltage 
\vg, of this parameter region. Because of the onset of mechanical 
self-excitation of the suspended nanotube \cite{strongcoupling, magdamping} (as 
e.g. already visible left to the electron number labels ``24'' and ``28'' in 
the figure) the measurement was restricted to comparatively low bias voltages. 
Satellite maxima accompanying the zero-bias Kondo anomaly at odd electron 
number can be recognized for a wide range of \nel. 

In Fig. \ref{shells} conductance traces $G(\vsd)$ taken at gate voltages 
corresponding to the center of a Coulomb blockade region are sorted by the 
filling state of shells corresponding to different longitudinal modes. In the 
odd valleys the distinct zero-bias peak can be observed. Additionally, 
satellite peaks systematically occur in all odd valleys with an overall 
increase of the Kondo temperature with increasing electron number. 
The universal behavior of the central peak and the non-universal behavior of 
the satellites is further illustrated also by Fig.~\ref{shellsrescaled}, where 
the conductance traces of Fig.~\ref{shells}(b) have been rescaled with the 
corresponding Kondo temperatures, similar to main text Fig.~1(d). In the even 
valleys $\nel=4N, \nel=4N+2$ the traces are nearly flat with weak features 
around zero bias whose origin is unclear. They gradually vanish with increasing 
magnetic field. 

It is evident that the level structure of our CNT sample is highly regular in 
terms of shell filling, allowing us to describe the fine structure of the Kondo 
resonances in different valleys by solely slightly adjusting the internal 
parameters \DSO\ and \DKK. We conclude that the fine structure of the Kondo 
effect occurs systematically in all valleys with odd electron numbers between 
11 and 27.

\bibliography{paper}

\begin{thebibliography}{57}
\bibitem [{foo()}]{footnote_PHS}%
  \BibitemOpen
  \href@noop {} {\ }\bibinfo {note} {The term particle-hole symmetry refers
  here to the fact that the corresponding operator $\hat{{\mathcal P}}$
  exchanges states at energies located symmetrically around a reference energy
  ($\varepsilon_d$ or $\varepsilon(\vec{B})$) within the quadruplet. It must
  not be confused with the particle-hole symmetry around the band
  gap.}\BibitemShut {Stop}%
\bibitem [{not()}]{note:G-asymm}%
  \BibitemOpen
  \href@noop {} {\ }\bibinfo {note} {The asymmetry of the peaks in the
  experiment is attributed to the difference in the coupling strength to the
  left and right contacts; for simplicity, in the theoretical curves the
  coupling to the left and right contacts is assumed to be of the same
  strength.}\BibitemShut {Stop}%
\bibitem [{exp()}]{expansion_point}%
  \BibitemOpen
  \href@noop {} {\ }\bibinfo {note} {Notice that the expansion point are in
  general function of temperature and bias. It turns out, however, that is
  sufficient to fix their value from conditions on the TDOS at zero temperature
  and bias. In fact for this choice of parameters the expansion points are of
  the order of the Kondo temperature, and hence do not play a role at high
  energies. However, their finite value is crucial in the Fermi liquid regime
  reached at low energies.}\BibitemShut {Stop}%
\end{thebibliography}

\end{document}